\documentclass[aps,prd,reprint,preprintnumbers,superscriptaddress,nofootinbib,floatfix]{revtex4-1}

\usepackage{amsmath,amssymb,amsfonts,dcolumn,color,graphicx,graphics,latexsym,placeins,epsfig}
\usepackage{epsfig}
\usepackage{bm}
\usepackage{slashed}
\usepackage{latexsym}
\usepackage{natbib}
\usepackage{url}
\usepackage{dcolumn}
\usepackage{color}
\usepackage{amsfonts,amssymb,amsmath}
\usepackage{graphicx,epsfig}
\usepackage{psfrag}
\usepackage{subfigure}
\usepackage{tabularx}
\usepackage{hyperref}
\hypersetup{colorlinks=true}
\usepackage{comment}

\newcommand{\be}{\begin{equation}}
\newcommand{\ee}{\end{equation}}
\newcommand{\ba}{\begin{eqnarray}}
\newcommand{\ea}{\end{eqnarray}}

\renewcommand{\inf}{\infty}

\begin{document}

\title{Searching for axions with quantum interferometry}
  
\preprint{IPPP/26/32}
  
\author{Tanmay Kumar Poddar}
\email{tanmay.k.poddar@durham.ac.uk}
\affiliation{Institute for Particle Physics Phenomenology (IPPP), Department of Physics, Durham University, Durham DH1 3LE, United Kingdom}

\author{Michael Spannowsky}
\email{michael.spannowsky@kit.edu}
\affiliation{Institute for Theoretical Physics, Karlsruhe Institute of Technology, 76131 Karlsruhe, Germany}
\affiliation{Institute for Quantum Materials and Technologies, Karlsruhe Institute of Technology, Karlsruhe 76131, Germany}

\begin{abstract}
Quantum phase measurements offer a complementary route to axion searches. We show that axion-photon interactions can imprint both Aharonov-Bohm (AB) and Berry phases in experimentally motivated quantum setups. For a coherently oscillating axion dark matter background, the induced effective current generates a time dependent magnetic flux in an rf-SQUID, leading to a measurable voltage signal through the Josephson phase. For representative benchmarks, this AB phase search reaches the minimum axion-photon coupling $g_{a\gamma\gamma}^{\mathrm{min}}\sim 7.8\times10^{-14}~\mathrm{GeV}^{-1}$ at axion mass $m_a\sim 10^{-10}~\mathrm{eV}$, with projected sensitivity that can improve on existing limits in that parameter space by roughly one to two orders of magnitude. We also identify a geometric phase observable in a Mach-Zehnder interferometer with an adiabatically rotating magnetic field, providing a proof-of-principle phase-based probe of meV-scale axions even when they do not constitute the dark matter, although sensitivity on the coupling remains weaker than current bounds with conservative tabletop benchmarks. Extending the analysis to a three level photon-axion quasiparticle (AQP)-axion system, with the AQP realized in a topological magnetic insulator, we find a potentially measurable THz Berry phase dominated by the AQP sector, furnishing a nontrivial validation of the formalism in a richer coupled system. These setups establish quantum phase observables as a useful new framework for axion searches, with immediate phenomenological promise in superconducting circuits and longer term potential in quantum enhanced interferometry.
\end{abstract}

\maketitle

\section{Introduction}
Pseudoscalar Nambu-Goldstone boson, such as the QCD (quantum chromodynamics) axion was initially proposed to dynamically solve the strong CP (charge conjugation-parity) problem, thereby explaining the smallness of the neutron electric dipole moment \cite{Baker:2006ts}. While such fields are produced massless at the level of spontaneous symmetry breaking, non-perturbative QCD instanton effects generate a small mass, thereby relating the axion mass to the symmetry breaking scale \cite{Peccei:1977hh,Weinberg:1977ma,Peccei:1977ur,Wilczek:1977pj,Adler:1969gk,Bell:1969ts,Peccei:2006as,Kim:2008hd}. More generally, string theory and higher-dimensional frameworks predict a broader class of axion-like particles (ALPs), which are typically ultralight and for which the mass and coupling scales are not necessarily related, as the explicit symmetry breaking scale does not need to coincide with the QCD confinement scale \cite{Preskill:1982cy,Abbott:1982af,Dine:1982ah,Svrcek:2006yi,Marsh:2015xka,Arvanitaki:2009fg}. Their weak couplings to Standard Model (SM) fields, in particular to photons, enable sensitive probes through precision measurements of photon propagation and polarization in astrophysical and laboratory settings \cite{AxionLimits,Arza:2026rsl}. Axion may also constitute a compelling dark matter (DM) candidate beyond the weakly interacting massive particle (WIMP) paradigm \cite{Bertone:2018krk,Chadha-Day:2021szb,OHare:2024nmr}.

The extremely weak coupling of axions to photons makes their detection challenging, motivating the development of precision experimental techniques. Cavity haloscopes such as ADMX \cite{Asztalos:2010,ADMX:2018gho,ADMX:2019uok,ADMX:2021nhd,ADMX:2024xbv,ADMX:2025vom,Crisosto:2019fcj,Stern:2016bbw} and HAYSTAC \cite{Brubaker:2016ktl,HAYSTAC:2018rwy,HAYSTAC:2020kwv,HAYSTAC:2023cam,HAYSTAC:2024jch} probe galactic axion DM in the $\mathcal{O}(1-20)~\mu\mathrm{eV}$ mass range via resonant conversion into microwave photons in a strong magnetic field. Complementary approaches such as the dielectric haloscope MADMAX \cite{MADMAX:2024sxs,Beurthey:2020yuq} enhance conversion through coherent emission across multiple interfaces, extending sensitivity to $\mathcal{O}(100)~\mu\mathrm{eV}$ masses.

At lower masses, the axion behaves as a coherently oscillating classical field with frequency set by its mass, rendering cavity-based techniques less effective due to the absence of resonant enhancement and increased low-frequency noise. This has led to the development of non-cavity approaches. Experiments such as ABRACADABRA~\cite{Ouellet:2018beu,Salemi:2021gck} probe ultralight axions in the $\mathcal{O}(0.3-8.3)~\mathrm{neV}$ range by measuring the induced oscillating magnetic field in a toroidal geometry, with future sensitivity extending to $\mathcal{O}(\mathrm{peV})$. Similarly, DMRadio~\cite{DMRadio:2022pkf} employs LC (inductor-capacitor) circuits to target masses from $\mathcal{O}(10^{-2}),\mathrm{neV}$ up to $\mathcal{O}(\mu\mathrm{eV})$. Additional proposals and experiments have further expanded the accessible parameter space~\cite{Lawson:2019brd,BREAD:2021tpx,QUAX:2024fut,Baryakhtar:2018doz,Adair:2022rtw,Quiskamp:2024oet,Alesini:2023qed,AxionLimits}.

Alternative strategies exploit engineered materials and condensed matter realizations of axion electrodynamics. Topological magnetic insulators can host axion quasiparticles, enabling sensitivity to $\mathcal{O}(\mathrm{meV})$ masses \cite{Marsh:2018dlj,SchutteEngel:2021,Cicoli:2026fqp}, while metamaterial-based setups can generate a tunable effective photon mass (plasma frequency), enhancing resonant axion-photon conversion for masses around $\mathcal{O}(10)~\mu\mathrm{eV}$ \cite{Kowitt:2023wob}.

The periodic axion background can induce geometric phases in quantum systems, including both Aharonov-Bohm (AB) \cite{Aharonov:1959fk} and Berry \cite{Berry:1984jv} phases. In \cite{Capolupo:2015cga}, the axion-induced Mukunda-Simon phase was studied, which is independent of adiabaticity and cyclic evolution and depends only on the path in Hilbert space. In \cite{Cao:2024lwg}, axions are treated as a background field modifying photon propagation, leading to birefringence and an associated geometric phase even in the absence of an external magnetic field.

Related effects have also been explored beyond axions. In particular, AB-type phases induced by hidden photons arise through kinetic mixing, which allows magnetic fields to penetrate nominally into the field free regions and enables experimental probes at the $\mathcal{O}(\mathrm{meV})$ scale~\cite{Arias:2016vxn}. Axion-induced AB phases can also lead to nontrivial topological configurations via modifications of the electromagnetic (EM) potential~\cite{Lambiase:2022rto}. Additional discussions of Berry and AB phases in related contexts can be found in \cite{Essin:2008rq,Chen:2024tsx,Badurina:2025idj,Ma:2026ujz}.

In this paper, we investigate an axion-induced AB phase arising in superconducting circuits. In an rf-SQUID (radio frequency-Superconducting Quantum Interference Device) \cite{Muck:2001,tinkham:2004} configuration, a Josephson junction (JJ) embedded in a superconducting loop provides a gauge-invariant phase difference that is directly sensitive to the magnetic flux threading the loop. This phase can be identified with an AB phase of the macroscopic superconducting wavefunction \cite{tinkham:2004}. In the presence of axion DM, the axion-photon interaction induces an effective current that modifies the EM field configuration, resulting in a time dependent shift in the magnetic flux. Consequently, the phase difference across the junction is altered, producing a measurable voltage signal. This provides a direct probe of axion-photon interactions using superconducting circuits.

In addition to the AB phase, we consider interferometric setups that probe a geometric (Berry) phase arising from axion-photon mixing in the presence of an adiabatically changing magnetic field. The mixing induces an additional phase beyond the conventional dynamical contribution, which can be detected as a shift in the interference pattern. Under adiabatic evolution, the coupled system remains in an instantaneous eigenstate, allowing the accumulation of a Berry phase determined by the geometry of a closed trajectory in parameter space. This space is spanned by the external magnetic field configuration, plasma properties (effective photon mass), photon frequency, and axion parameters such as its mass and coupling. The resulting phase is characterized by the solid angle enclosed by this trajectory, reflecting the geometric structure of the underlying axion-photon mixing Hamiltonian.

The paper is organized as follows. In section.~\ref{sec2}, we derive the axion-induced Aharonov-Bohm phase of the superconducting condensate and estimate the corresponding sensitivity to the axion-photon coupling using an rf-SQUID setup. In section.~\ref{sec3}, we compute the axion-induced Berry phase in a two-level axion-photon system and evaluate its detectability in interferometric experiments. We also analyze the Berry phase in a photon-graviton system for representative experimental configurations. In section.~\ref{sec4}, we extend the analysis to a three-level axion-axion quasiparticle (AQP)-photon system and outline a possible detection strategy. Finally, in section.~\ref{sec5}, we summarize and discuss our results.

We use natural system of units throughout the paper, unless stated otherwise.

\section{Axion-induced Aharonov-Bohm phase of the superconducting condensate}\label{sec2}

\begin{figure}[ht]
    \centering
\includegraphics[width=3in]{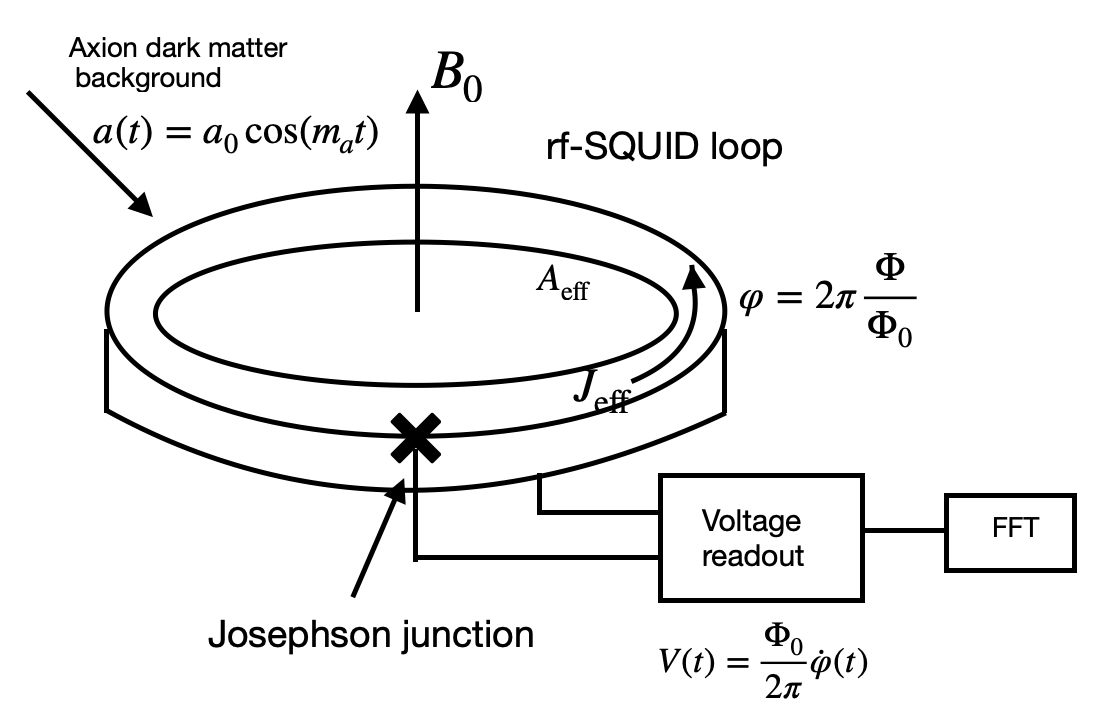}
   \caption{Schematic representation of detecting axion-induced Aharonov-Bohm phase using rf-SQUID loop}
    \label{plot1}
\end{figure}

In this section, we show that the axion field may induce an oscillatory magnetic flux in the superconducting pickup region via axion electrodynamics. The axion-photon interaction produces a gauge invariant phase shift of the superconducting condensate equivalent to a time dependent AB phase in the rf-SQUID. In an rf-SQUID, the magnetic flux threading the loop modulates the superconducting phase difference across the JJ. A time dependent flux therefore induces a time dependent phase, which, via the Josephson relation, generates a measurable voltage signal. The resulting voltage can be analyzed in the frequency domain using a fast Fourier transform (FFT) to identify the characteristic modulation induced by the axion field. A schematic representation of the detection of this axion-induced AB phase in the superconducting loop is shown in FIG.~\ref{plot1}.

We consider an axion field $a(x)$ with mass $m_a$ interacting with the EM field strength tensor $F_{\mu\nu}$ through the following Lagrangian \cite{Raffelt:1987im}
\begin{equation}
\begin{split}
\mathcal{L} =
-\frac{1}{4}F_{\mu\nu}F^{\mu\nu}
+\frac{1}{2}\partial_\mu a\,\partial^\mu a
-\frac{1}{2}m_a^2 a^2-\\
\frac{1}{4}g_{a\gamma\gamma}\, a\, F_{\mu\nu}\tilde F^{\mu\nu}- J^\mu A_\mu ,
\label{eq:L_axion}
\end{split}
\end{equation}
where $F_{\mu\nu}\tilde F^{\mu\nu} = -4\,\mathbf{E}\cdot \mathbf{B}$, with $\mathbf{E}$ and $\mathbf{B}$ are the background electric and magnetic fields, $J^\mu=(\rho,\mathbf{J})$ is the background four current density and $A_\mu$ is the four potential. Varying Eq.~\eqref{eq:L_axion} with respect to $A_\mu$ gives the inhomogeneous modified Maxwell's equations as
\begin{align}
\nabla\cdot \mathbf{E} &= \rho - g_{a\gamma\gamma}\, (\nabla a)\cdot \mathbf{B}, \label{eq:Gauss_ax}\\
\nabla\times \mathbf{B} - \partial_t \mathbf{E} &= \mathbf{J}
+ g_{a\gamma\gamma}\left( \dot a\, \mathbf{B} + \nabla a\times \mathbf{E}\right).
\label{eq:Ampere_ax}
\end{align}

Therefore, Eq.~\eqref{eq:Ampere_ax} shows that in the presence of a background magnetic field $\mathbf{B}=\mathbf{B}_0$ and a time oscillating axion background $a(t)$, one obtains an effective axion-induced current density
\begin{equation}
\mathbf{J}_{\rm eff} \simeq g_{a\gamma\gamma}\, \dot a(t)\, \mathbf{B}_0,
\label{eq:J_eff}
\end{equation}

where we consider that the axion field is spatially homogeneous over the apparatus i.e., $\nabla a\simeq 0$. In other words, the axion field is spatially uniform across the detector i.e., $k_aL\ll 1$, where $k_a$ is the axion DM momentum and $L$ is the size of the detector. Thus, the modified flux measured by the SQUID loop with loop area $S$ can be written as
\begin{equation}
\Phi(t)=\int _S \mathbf{B}\cdot d\mathbf{S}=\Phi_{\mathrm{ext}}+\delta\Phi_a(t),
\label{eq:1}
\end{equation}
where $\Phi_{\mathrm{ext}}$ is the externally applied magnetic flux threading the SQUID loop, and $\delta \Phi_a$ denotes the axion-induced flux. We use a standard magneto-quasistatic scenario, $\omega_a L\ll1$ ($\omega_a\simeq m_a$ is the axion oscillation frequency), where $\mathbf{J}_{\mathrm{eff}}$ can be recast as an effective current density that produces an extra flux. This also ensures that EM retardation effects are negligible and the induced flux follows the axion-driven current instantaneously.

We choose a pickup segment with mutual inductance $M_{\mathrm{eff}}$ such that the response due to the axion-induced flux is dominated by a localized region, and the flux in terms of axion-induced current $I_a(t)$ becomes
\begin{equation}
\delta\Phi_a(t)=M_{\mathrm{eff}}I_a, ~~~I_a(t)\equiv \int_\Sigma  \mathbf{J}_a\cdot d \mathbf{S}=g_{a\gamma\gamma}\dot{a}\int _\Sigma \mathbf{B}_0\cdot d\mathbf{S},   
\label{eq:2}
\end{equation}
where $\Sigma$ is the cross-section through which the effective current threads, which is set by the pickup geometry. The axion-induced current acts as an external source in the pickup region. Since the signal is time dependent, the superconducting condensate cannot fully screen it, and the response is effectively unsuppressed for frequencies above the inverse relaxation time. 

Thus, the axion-induced flux becomes
\begin{equation}
\delta \Phi_a=g_{a\gamma\gamma}\dot{a}M_{\mathrm{eff}} \Phi_{B_0}, ~~~    \Phi_{B_0}\equiv \int _\Sigma \mathbf{B}_0\cdot d\mathbf{S}\simeq B_0 A_{\mathrm{eff}}.
\label{eq:3}
\end{equation}
The superconducting pickup region denotes the localized volume where the axion-induced effective current is generated in the presence of the background magnetic field and $A_{\mathrm{eff}}$ is the effective area of the pickup region. The resulting magnetic flux is coupled to the SQUID loop through mutual inductance.

The standard Cooper pair condensate without any axion effect can be characterized as \cite{Ginzburg2009}
\begin{equation}
\psi=\sqrt{\rho}e^{i\theta},    
\label{tsm1}
\end{equation}
where $\rho$ is the number density of the Cooper pairs and $\theta$ is the quantum phase of the Cooper pair condensate wave function. Thus, the current  density for the superelectron in EM field due to the change of the gauge covariant derivative or the shift in the canonical momentum can be written as
\begin{equation}
J=\frac{iq}{2m}(\psi\nabla\psi^*-\psi^*\nabla\psi)-\frac{q^2}{m}\psi^* \mathbf{A}\psi,   
\label{tsm2}
\end{equation}
where $q=2e$ is the charge of the superelectron and $m$ is its mass. Using Eq.~\eqref{tsm1}, we obtain
\begin{equation}
\mathbf{J}=\frac{q\rho}{m}(\nabla\theta-q\mathbf{A}).   
\label{tsm3}
\end{equation}
The gauge-invariant phase gradient is $\nabla\theta-2e \mathbf{A}$. The single-valuedness around a closed loop gives the fluxoid quantization 
\begin{equation}
\oint(\nabla\theta-2e\mathbf{A})\cdot d\mathbf{l}=2\pi n, ~~~   n=0,\pm1,\pm2,\ldots 
\end{equation}
where $n$ denotes the winding number, a topological integer that counts the number of times the phase of the complex order parameter winds by $2\pi$ around a closed loop. For a superconducting loop containing one JJ, the gauge-invariant phase quantization can be written as
\begin{equation}
\varphi+\frac{2\pi\Phi}{\Phi_0}=2\pi n, 
\end{equation}
where $\Phi_0=2\pi/2e$. In our analysis, we consider the lowest fluxoid state corresponding to $n=0$. Therefore, the absolute value of the phase difference across a JJ is obtained as
\begin{equation}
\varphi=2\pi\frac{\Phi}{\Phi_0}, 
\label{eq:5}
\end{equation}
where $\varphi$ is the AB phase, implies a charged quantum wavefunction acquires phase proportional to enclosed magnetic flux. The axion effective current density changes the flux in the SQUID loop which yields additional contribution to the Josephson phase as
\begin{equation}
\delta\varphi_a(t)=\frac{2\pi}{\Phi_0}\delta\Phi_a(t).    
\label{eq:6}
\end{equation}

For the readout, we write the Josephson equation as 
\begin{equation}
V(t)=\frac{\Phi_0}{2\pi}\dot{\varphi}(t)    
\label{eq:7}
\end{equation}
where $\dot{\varphi}$ is obtained from the fluxoid relation for an rf-SQUID loop, given in Eq.~\eqref{eq:5} with the total flux $\Phi(t)=\Phi_{\mathrm{ext}}+\mathcal{L} I+\delta\Phi_a(t)$ and $\mathcal{L}$ is the inductance of the superconducting loop. Therefore, the axionic contribution to the voltage readout becomes
\begin{equation}
\delta V_a(t)=\frac{d}{dt}\delta\Phi_a(t), 
\label{eq:8}
\end{equation}

where the SQUID output voltage is directly proportional to the time-varying flux. In our analysis, the axion field behaves as a time-oscillating DM with oscillation frequency $\omega_a=m_a$ given as
\begin{equation}
a(t)=\frac{\sqrt{2\rho_{\mathrm{DM}}}}{m_a}\cos(m_a t),    
\label{eq:9}
\end{equation}
where $\rho_{\mathrm{DM}}$ denotes the energy density of the DM in the Universe. Therefore, the amplitude of the axion-induced DM flux is obtained from Eq.~\eqref{eq:3} as
\begin{equation}
\delta\Phi_a^{0}=g_{a\gamma\gamma}\sqrt{2\rho_{\mathrm{DM}}}M_{\mathrm{eff}}\Phi_{B_0},    
\label{eq:10}
\end{equation}
where the signal amplitude is independent of axion DM mass.

To estimate the sensitivity of the axion-photon coupling, we consider the SQUID has the flux noise spectral density $S_\Phi^{1/2}$ (in $\Phi_0/\sqrt{\mathrm{Hz}}$ or $\mathrm{Wb}/\sqrt{\mathrm{\mathrm{Hz}}}$). For a narrowband axion signal at $\omega_a=m_a$, the detectable flux amplitude scales like 
\begin{equation}
\delta\Phi_a^{\mathrm{min}}\simeq \frac{S_\Phi^{1/2}}{\sqrt{T_{\mathrm{eff}}}}, ~~~T_{\mathrm{eff}}\equiv \mathrm{min}~(T, \tau_c),
\label{eq:11}
\end{equation}
where $T$ is the integration time $\tau_c\sim 2\pi/(m_av^2)$ is the axion coherence time with $v$ denotes the DM velocity. Thus, the sensitivity to the axion-photon coupling can be obtained as

\begin{equation}
g_{a\gamma\gamma}^{\mathrm{min}}\simeq \frac{S^{1/2}_\Phi}{\sqrt{T_{\mathrm{eff}}}\sqrt{2\rho_{\mathrm{DM}}}M_{\mathrm{eff}}\Phi_{B_0}},    
\label{eq:12}
\end{equation}

which follows directly by equating the axion-induced flux amplitude in Eq.~\eqref{eq:10} to the minimum detectable flux set by the SQUID noise in Eq.~\eqref{eq:11}, and solving for the smallest detectable axion-photon coupling. To estimate the sensitivity to the axion-photon coupling, we adopt realistic benchmark parameters. We take the effective inductance to be $ M_{\mathrm{eff}} = 50~\mathrm{nH}$ and the background magnetic flux $\Phi_{B_0} = 5~\mathrm{T}\cdot 5~\mathrm{cm^2}$. The local DM density is assumed to be $\rho_{\mathrm{DM}} = 0.3~\mathrm{GeV~cm^{-3}}$ \cite{Read:2014qva}.

The minimum detectable flux is given by $\delta\Phi^{\mathrm{min}}=S_\Phi^{1/2}/\sqrt{T_{\mathrm{eff}}}$, where we use an effective integration time $T_{\mathrm{eff}} = 6~\mathrm{s}$ and a flux noise spectral density $ S_\Phi^{1/2}=0.1~\mu\Phi_0/\sqrt{\mathrm{Hz}}$, with the superconducting flux quantum $\Phi_0=\pi/e$. Numerically, $\Phi_0 = 2.07\times 10^{-15}~\mathrm{Wb}$, $1~\mathrm{H}=1~\mathrm{Wb/A}=4.1\times10^{21}~\mathrm{GeV^{-1}}$, and $1~\mathrm{Wb}=5.05\times10^{15}$. Hence, we obtain the sensitivity on $g_{a\gamma\gamma}$ from Eq.~\eqref{eq:12} as
\begin{widetext}
\begin{equation}\label{sensei}
g_{a\gamma\gamma}^{\mathrm{min}}\simeq 7.8\times 10^{-14}~\mathrm{GeV}^{-1} \Big(\frac{S^{1/2}_\Phi}{0.1~\mu\Phi_0/\sqrt{\mathrm{Hz}}}\Big) \Big(\frac{6~\mathrm{s}}{T_{\mathrm{eff}}}\Big)^{1/2} \Big(\frac{0.3~\mathrm{GeV}/\mathrm{cm}^{3}}{\rho_{\mathrm{DM}}}\Big)^{1/2} \Big(\frac{50~\mathrm{nH}}{M_{\mathrm{eff}}}\Big)\Big(\frac{5~\mathrm{T}. 5~\mathrm{cm^2}}{\Phi_{B_0}}\Big).
\end{equation}
\end{widetext}

Equations.~\eqref{eq:12} and ~\eqref{sensei} are written at the level of effective coupled input flux compared against the input-referred flux noise. Thus, the omitted circuit details should only modify the overall transfer function normalization, at most by an order one factor, without changing the basic sensitivity scaling. The rf-SQUID setup considered here functions as an effectively nonresonant broadband sensor for an axion DM signal. For a given axion mass, however, the signal itself is narrowband, appearing as a quasi-monochromatic spectral line at frequency $\omega \simeq m_a$ with fractional linewidth $\Delta \omega/\omega \sim v^2 \sim 10^{-6}$. In practice, this corresponds to a very narrow peak in the FFT of the output voltage. As a result, the projected sensitivity is approximately independent of $m_a$ when the axion coherence time exceeds the observation time, $\tau_c > T$. For larger masses, where the coherence time becomes shorter than the integration time, $\tau_c < T$, the sensitivity degrades as $g_{a\gamma\gamma}^{\min}\propto m_a^{1/2}$. This behavior corresponds to the idealized limit in which both the flux-noise spectral density and the detector response are frequency independent.

To model the detector more realistically, we parametrize the frequency response of the pickup and readout chain by a single-pole transfer function, $R(\omega)=[1+(\omega/\omega_c)^2]^{-1/2}$, where $\omega_c$ denotes the effective bandwidth scale. This form reproduces the ideal broadband limit $R(\omega)\simeq 1$ for $\omega \ll \omega_c$, while capturing the expected high frequency roll-off for $\omega \gg \omega_c$. As a representative conservative benchmark, we take $f_c=\omega_c/2\pi \sim 10~\mathrm{MHz}$, appropriate to a standard flux-locked-loop SQUID readout \cite{Drung:2005}.

The flux-noise power spectral density may also exhibit nontrivial frequency dependence. We model the corresponding amplitude spectral density as
$S_\Phi^{1/2}(f)=S^{1/2}_{\Phi,\omega}[1+(f_{1/2}/f)]^{1/2}$ where $S_{\Phi,w}^{1/2}$ is the white-noise floor and $f_{1/f}$ is the low frequency corner. This parametrization incorporates the standard low frequency $1/f$-type excess noise characteristic of SQUIDs and related superconducting circuits \cite{Koch:2007jdf}. As a conservative choice, we take $S_{\Phi,w}^{1/2}=0.1~\mu\Phi_0/\sqrt{\mathrm{Hz}}$ and $f_{1/f}=1~\mathrm{Hz}$ \cite{OUKHANSKI:2002}.

Therefore, the rf-SQUID setup provides broadband coverage in the sense that it can probe a wide range of axion masses without requiring resonant retuning. At the same time, for each fixed $m_a$, the axion DM signal remains intrinsically narrowband. The resulting sensitivity curve is thus broadband in coverage, while the signal at each mass point is a narrow spectral feature.

\begin{figure*}[ht]
    \centering
\includegraphics[width=5in]{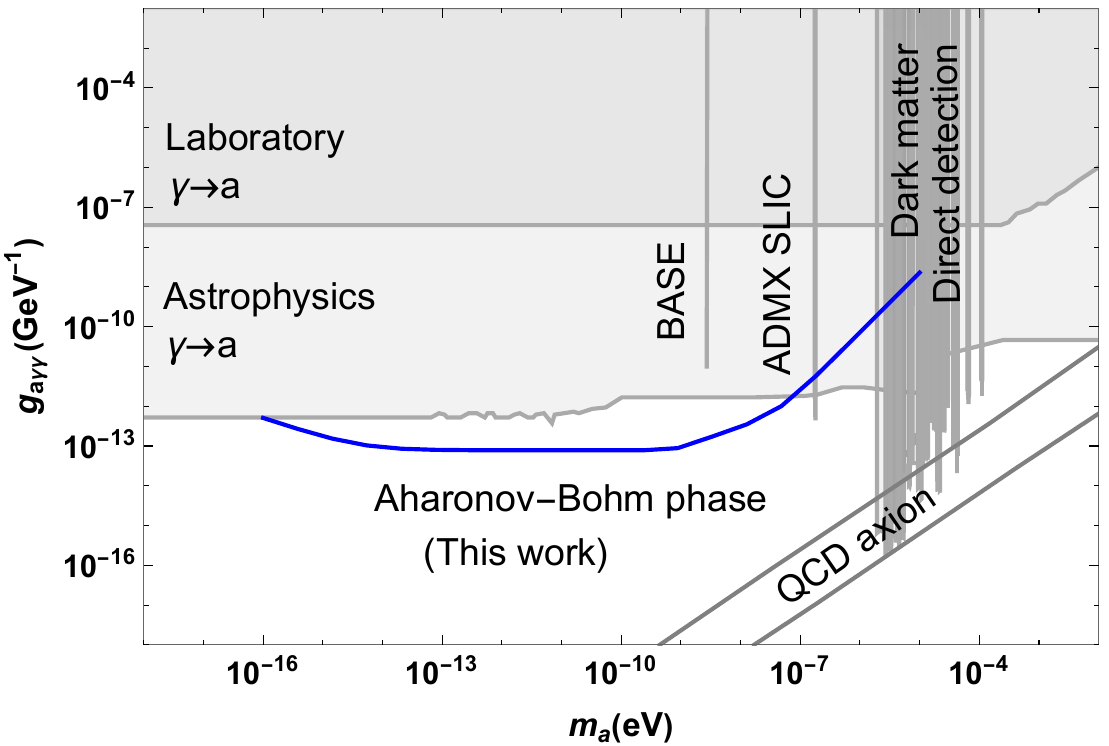}
    \caption{Sensitivity to the axion-photon coupling from the Aharonov-Bohm phase measurement using rf-SQUID loop.}
    \label{plotsensitivity}
\end{figure*}

In FIG.~\ref{plotsensitivity}, we present the projected sensitivity to the axion-photon coupling $g_{a\gamma\gamma}$ as a function of the axion mass, obtained from the measurement of the axion-induced AB phase with an rf-SQUID loop. The projected reach is shown together with existing bounds from laboratory experiments \cite{AxionLimits}, astrophysical observations \cite{AxionLimits}, and DM direct detection searches, including ADMX \cite{AxionLimits},
QUAX \cite{Alesini:2019ajt,Alesini:2020vny,Alesini:2022lnp,QUAX:2023gop,QUAX:2024fut}, RBF+UF \cite{DePanfilis:1987,Wuensch:1989sa,PhysRevD.42.1297,Hagmann:1996qd}, ORGAN \cite{McAllister:2017lkb,Quiskamp:2022pks,Quiskamp:2023ehr,Quiskamp:2024oet}, ADMX SLIC \cite{Crisosto:2019fcj}, and BASE \cite{Devlin:2021fpq}. The QCD axion band is also shown \cite{AxionLimits}. For $m_a\lesssim 4\times 10^{-15}~\mathrm{eV}$, the sensitivity degrades due to the low frequency $1/f$ excess noise, assuming a knee frequency $f_{1/f}=1~\mathrm{Hz}$. At larger masses, the reach weakens once the axion coherence time becomes shorter than the integration time, which occurs around $m_a\simeq 7\times 10^{-10}~\mathrm{eV}$. In addition, the sensitivity is further suppressed at $m_a\simeq 4\times 10^{-8}~\mathrm{eV}$ due to the finite detector bandwidth, corresponding to a response cutoff frequency $f_c=10~\mathrm{MHz}$. The projected sensitivity improves upon the currently existing bounds by approximately one to two orders of magnitude.

\section{Axion-induced Berry phase from photon interferometry}\label{sec3}

\begin{figure}[ht]
    \centering
\includegraphics[width=3in]{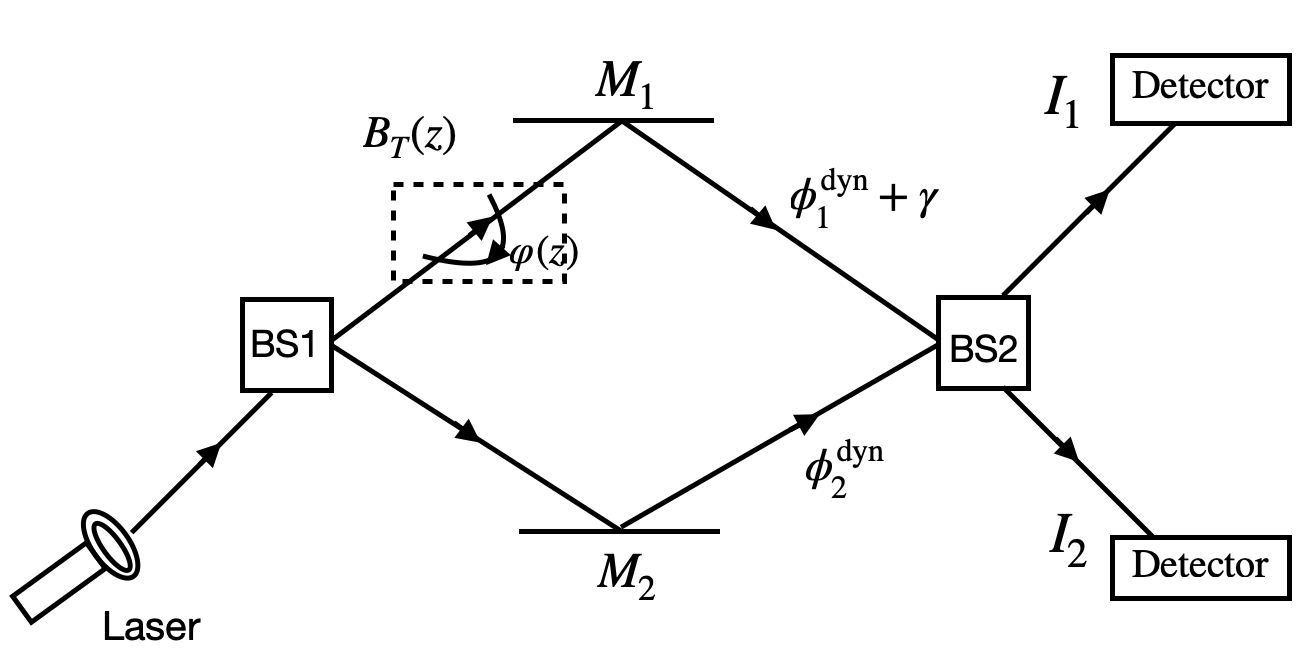}
   \caption{Schematic representation of detecting axion-induced Berry phase using Mach-Zehnder interferometer}
    \label{plot2}
\end{figure}

The presence of an external magnetic field can induce mixing between photons and ALPs, allowing the axion-photon system to exhibit a geometric (Berry) phase during the adiabatic evolution of the magnetic field. In the following, we consider a two level axion-photon system and study the evolution of the particle states as they propagate through a magnetic field over a distance.

A schematic illustration of the proposed setup for measuring the axion-induced Berry phase in an axion-photon system using a Mach-Zehnder interferometer (MZI) \cite{Zehnder1891,Mach1892} is shown in FIG.~\ref{plot2}. The interferometer is fed by a coherent laser source, which is split into two optical paths by the first beam splitter (BS1). One arm contains an adiabatically varying transverse magnetic field, $B_T(z)$, which induces axion-photon mixing, while the second arm serves as a reference and is free from any external field. The beams propagating along the two arms are reflected by mirrors $M_1$ and $M_2$, and subsequently recombined at the second beam splitter (BS2). The resulting interference pattern, recorded at the detector, encodes both the dynamical phase and the geometric (Berry) phase acquired during propagation. To isolate the geometric contribution, a control measurement can be performed in the absence of the magnetic field, in which case only the dynamical phase is present. The Berry phase can then be extracted from the differential fringe shift between the two configurations.

In the presence of a static transverse magnetic field $B_T$, photons can mix with ALPs. Consider a monochromatic laser beam light with frequency $\omega$ propagating along the $z$ direction. Choosing the polarization basis such that the photon polarization perpendicular to the magnetic field decouples, only the parallel polarization mode $\gamma_{||}$ mixes with the axion field $a$. The evolution of the axion-photon system can then be described by the following Schr\"odinger-like equation along the propagation direction as \cite{Raffelt:1987im}
\begin{equation}
i\partial_z
\begin{pmatrix}
\gamma_\parallel(z)\\
a(z)
\end{pmatrix}
=
\begin{pmatrix}
\Delta_\gamma & \Delta_{a\gamma}\\
\Delta_{a\gamma} & \Delta_a
\end{pmatrix}
\begin{pmatrix}
\gamma_\parallel(z)\\
a(z)
\end{pmatrix},
\label{berry1}
\end{equation}
where the mixing parameters are given by 
\begin{equation}
\Delta_{a\gamma} = \frac{1}{2}g_{a\gamma\gamma} B_T, \qquad
\Delta_a = -\frac{m_a^2}{2\omega}, \qquad
\Delta_\gamma = -\frac{\omega_{\rm pl}^2}{2\omega}.
\label{berry2}
\end{equation}
Here, $\omega_{\mathrm{pl}}$ denotes the plasma frequency, which effectively endows the photon with a mass in a medium. In vacuum, however, $\omega_{\mathrm{pl}}=0$ and therefore $\Delta_{\gamma}=0$. 

The interaction eigenstates $(\gamma_{||}~~a)^{T}$ are not the propagation eigenstates of the system. To determine the true eigenmodes, the Hamiltonian matrix in Eq.~\eqref{berry1} must be diagonalized. This can be achieved by introducing a rotation characterized by the mixing angle $\theta$, relating the propagation eigenstates $(\gamma_1~~\gamma_2)^{T}$ to the interaction eigenstate through
\begin{equation}
\begin{pmatrix}
\gamma_1\\
\gamma_2
\end{pmatrix}
=
\begin{pmatrix}
\cos\theta & \sin\theta\\
-\sin\theta & \cos\theta
\end{pmatrix}
\begin{pmatrix}
\gamma_\parallel\\
a
\end{pmatrix}.
\label{berry3}
\end{equation}

The mixing angle $\theta$ is chosen such that the Hamiltonian becomes diagonal, yielding
\begin{equation}
\qquad
\tan 2\theta = \frac{2\Delta_{a\gamma}}{\Delta_\gamma-\Delta_a}.
\label{berry4}
\end{equation}
In the rotated basis, the propagation eigenstates $\gamma_1$ and $\gamma_2$ evolve independently with eigenvalues
\begin{equation}
\lambda_\pm = \frac{\Delta_a+\Delta_\gamma}{2} \pm \frac{1}{2}\sqrt{(\Delta_\gamma-\Delta_a)^2 + 4\Delta_{a\gamma}^2 }.
\label{berry5}
\end{equation}

A simple way to generate such a geometric phase in the axion-photon system is to rotate the direction of the transverse magnetic field along the propagation direction, adiabatically. In this case, the mixing term acquires a controllable complex phase. Concretely, consider a transverse magnetic field $B_T(z)$ whose direction in the transverse plane is characterized by an azimuthal angle $\varphi(z)$. In an appropriate polarization basis, the mixing term generalizes as
\begin{equation}
\Delta_{a\gamma}\to\Delta_{a\gamma} e^{-i\varphi(z)}.  
\end{equation}
A convenient way to describe the axion-photon system is through the Bloch-vector representation of the Hamiltonian
\begin{equation}
H=
\begin{pmatrix}
\Delta_\gamma & \Delta_{a\gamma}e^{-i\varphi}\\
\Delta_{a\gamma}e^{i\varphi} & \Delta_a
\end{pmatrix},    
\label{am1}
\end{equation}

which can be written as
\begin{equation}
\begin{aligned}
H &= \frac{\Delta_a+\Delta_\gamma}{2}\, \mathbb{I}
+\frac{1}{2}\,\boldsymbol{\Delta}\cdot \boldsymbol{\sigma},\\
\boldsymbol{\Delta} &=
\left( 2\Delta_{a\gamma}\cos\varphi,\, 2\Delta_{a\gamma}\sin\varphi,\, \Delta_\gamma-\Delta_a\right),
\label{am2}
\end{aligned}
\end{equation}

where $\mathbb{I}$ denotes the identity matrix and $\boldsymbol{\sigma}$ represents the vector of Pauli matrices. In this form, the interaction Hamiltonian is formally equivalent to that of a spin-$1/2$ particle in an effective magnetic field $\boldsymbol{\Delta}$. The unit vector $\hat{\Delta}={\boldsymbol{\Delta}}/|\boldsymbol{\Delta}|$,  
therefore defines a point on the Bloch sphere, so that the unit vector $\hat{\Delta}$ sweeps a closed trajectory as $\varphi$ evolves from $0$ to $2\pi$. If the system parameters (e.g. magnetic field) vary sufficiently slowly, the instantaneous eigenstates follow the direction of $\hat{\Delta}$ adiabatically. When $\hat{\Delta}$ traces a closed loop on the Bloch sphere, the eigenstates acquire a geometric (Berry) phase given by
\begin{equation}
\gamma_{\pm}=\mp \frac{1}{2}\Omega [\boldsymbol{\Delta}],    
\label{berry7}
\end{equation}
where $\Omega [\boldsymbol{\Delta}]$ is the solid angle enclosed by the trajectory of $\hat{\Delta}$. The exact sign of the Berry phase depends on the choice of the eigenstate labeling and the orientation of the closed path. 

Since the solid angle satisfies $0\leq \Omega \leq 4\pi$, the magnitude of the Berry phase is bounded by $|\gamma|\leq 2\pi$. Consequently, the Berry phase cannot grow indefinitely. Importantly, this phase is purely geometric, as it depends only on the area enclosed on the Bloch sphere and not on the rate at which the parameters evolve.

When the transverse magnetic field direction varies along the propagation path, the locally defined photon polarization basis becomes $z$-dependent, inducing a coupling to the orthogonal mode proportional to $\partial_z\varphi$. In the adiabatic regime $|\partial_z\varphi| \ll \Delta_{\rm osc}=\sqrt{(\Delta_\gamma-\Delta_a)^2+4\Delta_{a\gamma}^2}$, this coupling is parametrically suppressed, and the orthogonal polarization $(\gamma_\perp)$ remains negligibly populated, allowing a consistent reduction to the effective two-level $(\gamma_\parallel,a)$ system. In particular, one can show $\gamma_\perp\sim (\varphi^\prime/\Delta_{\mathrm{osc}})\gamma_\parallel$ and for $\varphi^\prime/\Delta_{\mathrm{osc}}\ll1$, $\gamma_\perp$ is decoupled from the effective $2\times 2$ Hamiltonian.

Comparing Eq.~\eqref{am2} with a generic vector in spherical polar coordinates 
\begin{equation}
\boldsymbol{\Delta}= |\boldsymbol{\Delta}|(\sin\vartheta\cos\varphi, \sin\vartheta\sin\varphi, \cos\vartheta),
\label{am3}
\end{equation}
where the direction of $\boldsymbol{\Delta}$ on the Bloch sphere is specified by the polar angle $\vartheta$ and azimuthal angle $\varphi$, we obtain the transverse and longitudinal components of $\boldsymbol{\Delta}$ as
\begin{equation}
\Delta_{\perp}=\sqrt{\Delta_x^2+\Delta^2_y}=2\Delta_{a\gamma},~~~~\mathrm{and,}~~~~  \Delta_z=\Delta_\gamma-\Delta_a,
\label{am4}
\end{equation}
respectively. We can also also define the polar angle on the Bloch sphere as 
\begin{equation}
\tan\vartheta=\frac{\Delta_{\perp}}{\Delta_z}=\frac{2\Delta_{a\gamma}}{\Delta_\gamma-\Delta_a}.
\label{am5}    
\end{equation}

 The resulting Berry phase depends on the parameters $(g_{a\gamma\gamma}, B_T, m_a, \omega_{\mathrm{pl}} )$ through the mixing angle $\theta$. On the Bloch sphere, the polar angle of the vector $\Delta$ is given by $\vartheta=2\theta$, while the azimuthal angle corresponds to $\varphi$. Thus, as $\varphi$ varies from $0$ to $2\pi$ with fixed $\vartheta$, the vector $\hat{\Delta}$ traces a circle of latitude on the Bloch sphere. The solid angle enclosed by this loop is $\Omega=2\pi(1-\cos\vartheta)$, which yields the Berry phase
\begin{equation}
\begin{split}
\gamma_{\pm}=\mp 2\pi \sin^2\theta=\mp \pi\Bigg(1-\frac{\Delta_\gamma-\Delta_a}{\sqrt{(\Delta_\gamma-\Delta_a)^2+4\Delta^2_{a\gamma}}}\Bigg)\\~~\mathrm{(mod~2\pi)}.   
\end{split}
\label{berry8}
\end{equation}

In the regime of small mixing, $|\theta|\ll1$, Eq.~\eqref{berry8} reduces to
\begin{equation}
\gamma_{\pm}\simeq \mp 2\pi \Big(\frac{\Delta_{a\gamma}}{\Delta_\gamma-\Delta_a}\Big)^2=\mp 2\pi \Big(\frac{g_{a\gamma\gamma}\omega B_T}{m^2_a-\omega^2_{\mathrm{pl}}}\Big)^2,~~\mathrm{(mod~2\pi)}. 
\label{berry9}
\end{equation}
Equation.~\ref{berry9} represents the axion-induced Berry phase in the small mixing limit, which vanishes in the absence of axion-photon coupling, $g_{a\gamma\gamma}\to 0$ or magnetic field, $B_T\to 0$. In absence of plasma, $\omega_{\mathrm{pl}}\to 0$, and Eq.~\ref{berry9} suggests that very light axions could lead to a large geometric phase. However, the Berry phase is bounded by $|\gamma|\leq 2\pi$, and the small-mixing approximation breaks down once the mixing becomes large. In contrast, for large axion masses the Berry phase decreases rapidly, scaling as $\gamma_{\pm}\propto 1/m^4_a$.

The Berry phase manifests itself as an observable phase difference in an interferometric setup. Importantly, the phase shift is purely geometric. The Berry phase of a single state cannot be measured directly. Instead, an interferometer measures the relative phase difference between two optical paths. It depends only on the closed trajectory traced by the mixing parameters in parameter space and is independent of the propagation length. In comparison, the axion-photon oscillation probability varies as square of the oscillation length.

Consider an interferometer with two arms. A beam-splitter is used to split the laser beam into two optical paths. In one arm, photons propagate through a region where the transverse magnetic field direction rotates slowly along the propagation path, leading to the accumulation of a Berry phase. In the other arm, photons propagate through a region with no magnetic field and therefore do not experience such geometric phase accumulation. When the two beams recombine through another beam-splitter, their superposition produces interference fringes whose positions are shifted due to the additional geometric phase, which can be measured through a photo-detector. 

 The geometric phase can therefore be extracted experimentally by subtracting the dynamical contribution, which can be determined by performing the same measurement in the absence of magnetic field rotation. In that case, only the dynamical phase $\Delta\phi_{\mathrm{dyn}}$ contributes to the interference pattern.

Furthermore, since only the parallel polarization mode $\gamma_{\parallel}$ mixes with the axion field and acquires the Berry phase, the axion effect generally changes the polarization state of the light. This typically leads to ellipticity, and only in special cases can it be described as a pure rotation of the polarization angle.

To estimate the sensitivity to the axion-photon coupling, we consider that the Berry phase manifests itself as a measurable phase shift in the intensity output of an interferometer. At first glance, the sensitivity appears to extend to arbitrarily small axion masses because the Berry phase $\gamma_{\pm}$ increases as $m_a$ decreases (see Eq.~\eqref{berry9}). However, the derivation is valid only in the small-mixing regime, $|\theta|\ll1$, which requires
\begin{equation}
\Bigg|\frac{\Delta_{a\gamma}}{\Delta_\gamma-\Delta_a}\Bigg|\ll1,
\qquad
m_a \gg \sqrt{g_{a\gamma} B_T \omega},
\end{equation}
in vacuum.

In an interferometric measurement, the observable quantity is the photon number detected at the output. If the average number of photons detected over an integration time $T$ is $N$, the statistical fluctuation is $\Delta N=\sqrt{N}$. This fluctuation corresponds to shot noise, which translates into a phase uncertainty $\delta\phi_{\mathrm{shot}}=1/\sqrt{N}$, representing the standard quantum limit (SQL).

To estimate the achievable phase sensitivity, we consider a monochromatic laser beam with wavelength $\lambda=1064~\mathrm{nm}$, corresponding to the fundamental transition wavelength of a Nd:YAG laser, which provides excellent frequency stability. This wavelength corresponds to a photon energy $\omega\simeq 1.2~\mathrm{eV}$. For a magnetic field strength $B=10~\mathrm{T}$, optical power $P=1~\mathrm{W}$, and integration time $T=10^4~\mathrm{s}\simeq 2.7~\mathrm{h}$, the total number of detected photons is

\begin{equation}
N=\frac{PT}{\omega}\simeq 5.2\times10^{22},
\qquad
\delta\phi_{\mathrm{shot}}\simeq4.4\times10^{-12}.
\end{equation}

Therefore, in order to observe the Berry phase signal, the accumulated phase $\gamma_{\pm}$ must exceed the quantum phase noise limit $\delta\phi_{\mathrm{shot}}$, i.e., $\gamma_{\pm}\gtrsim \delta\phi_{\mathrm{shot}}$. Although the shot-noise limit suggests a phase sensitivity as small as $\mathcal{O}(10^{-12})$, realistic interferometric measurements are typically limited by technical noise sources such as laser frequency fluctuations, mechanical vibrations, and thermal drifts. These effects set a more practical phase sensitivity around $\delta\phi_c\sim\mathcal{O}(10^{-6})$ \cite{Dandridge1982,Hariharan2003,Cronin2009}.

\subsubsection*{Adiabaticity condition}
We write the Schr\"odinger equation for the propagation along $z-$direction as $i\partial_z\Psi(z)=H(z)\Psi(z)$. The state vector in the basis of instantaneous eigenstates of the axion-photon Hamiltonian $H(z)|n(z)\rangle=\lambda_n(z)|n(z)\rangle$ can be written as
\begin{equation}
|\Psi(z)\rangle=\sum_n c_n(z) e^{-i\int ^z_{z_0}\lambda_n(z^\prime)dz^\prime} |n(z)\rangle. 
\label{cm1}
\end{equation}
Therefore, using Eq.~\ref{cm1} in the evolution equation, we obtain
\begin{equation}
\dot{c}_n(z)=-\sum_{m\neq n} c_m(z)\langle n(z)|\partial_z m(z)\rangle e^{-i\int ^z_{z_0}(\lambda_m-\lambda_n)dz^\prime},  
\label{cm2}
\end{equation}
where the dot denotes the differentiation with respect to $z$. The adiabaticity condition requires that the off-diagonal transition amplitudes are negligible, i.e., $|\langle n|\partial_z m\rangle|\ll |\lambda_n-\lambda_m|$. For the axion-photon two level system, we can write from Eq.~\ref{berry5}
\begin{equation}
\lambda_+-\lambda_-=\Delta_{\mathrm{osc}}=\sqrt{(\Delta_\gamma-\Delta_a)^2+4\Delta_{a\gamma}^2}.
\label{cm3}
\end{equation}
Since the $z-$dependence of the Hamiltonian arises only through the azimuthal angle $\varphi(z)$ of the transverse magnetic field, we can write $\langle+|\partial_z-\rangle\simeq (1/2)\sin(2\theta)(d\varphi/dz)$, where $|+\rangle$ and $|-\rangle$ denote the instantaneous propagation eigenstates corresponding to the eigenvalues $\lambda_+$ and $\lambda_-$, respectively. Hence, the adiabaticity condition becomes
\begin{equation}
\Big|\frac{d\varphi}{dz}\Big|\ll \Delta_{\mathrm{osc}}, 
\label{cm4}
\end{equation}
up to $\mathcal{O}(1)$ factor fixed by the mixing angle. We can write parametrically $L_{\mathrm{osc}}\sim \Delta^{-1}_{\mathrm{osc}}$ and if one full $2\pi$ rotation of the field is completed over a length $L_{\mathrm{rot}}$ then $|d\varphi/dz|\sim 2\pi/L_{\mathrm{rot}}$, such that the adiabaticity condition can be interpreted as $L_{\mathrm{rot}}\gg L_{\mathrm{osc}}$, which implies that the rate of the magnetic field rotation should be slow enough that the system remains in the same instantaneous eigenstates throughout propagation. In the configuration considered here, the magnitude of the transverse magnetic field and the diagonal entries are taken to be $z-$independent, so the mixing angle $\theta$ is constant and the only source of $z$-dependence is the azimuthal angle $\varphi(z)$.

The combined requirements of small mixing and adiabaticity results $m_a^2\gg 4\pi\omega/L_{\mathrm{rot}}$. For the benchmark values $\omega\sim 1.2~\mathrm{eV}$ and $L_{\mathrm{rot}}\sim 1~\mathrm{m}$, we obtain $m_a\gg 1.7~\mathrm{meV}$. Axions with masses below $\mathcal{O}(~\mathrm{meV})$ can also be probed in principle, provided the magnetic field rotation length satisfies $L_{\mathrm{rot}}\gg 1~\mathrm{m}$, although realizing such a large scale is challenging in a tabletop setup. 

Therefore, conservative choices of the input parameters leads to the following estimate for the minimum detectable axion-photon coupling
\begin{widetext}
\begin{equation}\label{fre}
g_{a\gamma\gamma}^{\mathrm{min}}
\simeq
6.8\times10^{-4}~\mathrm{GeV}^{-1}
\left(\frac{m_a}{2~\mathrm{meV}}\right)^2
\left(\frac{1.2~\mathrm{eV}}{\omega}\right)
\left(\frac{10~\mathrm{T}}{B_T}\right)
\left(\frac{\delta\phi_{\mathrm{c}}}{10^{-6}}\right)^{1/2}.
\end{equation}
\end{widetext}
The phase readout constitutes a non-resonant phase-shift search across axion parameter space, rather than a narrowband line search. The lower end of this interval is set by the requirement that the small-mixing approximation and the adiabaticity condition remain valid, while at larger masses the sensitivity to the axion-photon coupling rapidly degrades. We emphasize that this interferometric search for an axion-induced Berry phase does not rely on axions constituting the DM.

For axions with masses of $\mathcal{O}(\mathrm{meV})$, the current bound on the axion-photon coupling is $g_{a\gamma\gamma}\lesssim 5\times10^{-11}~\mathrm{GeV}^{-1}$, as inferred from globular cluster observations \cite{Ayala:2014pea,Dolan:2022kul}. Consequently, the sensitivity obtained in Eq.~\eqref{fre} from the Berry phase measurement remains weaker than the existing limit. In principle, the reach can be improved by increasing the external magnetic field, the photon energy, the magnetic field rotation length, or more generally by enhancing the phase sensitivity of the setup. In a realistic tabletop experiment, however, such modifications may jeopardize the validity of the small mixing and adiabatic approximations, while also introducing additional noise and decoherence. An idealized possibility is to employ an entangled $N$ photon state, such that all photons acquire the same geometric phase coherently. In that case, the phase uncertainty scales as $\delta\phi \sim 1/N$, corresponding to the Heisenberg limit (HL) \cite{Pezze:2005zup}. Under this idealized assumption, the minimum reachable coupling could be as low as $g_{a\gamma\gamma}^{\min}\sim 3\times10^{-12}~\mathrm{GeV}^{-1}$, improving upon the current bound for $\mathcal{O}(\mathrm{meV})$ axions by approximately one order of magnitude.

\subsubsection*{Berry phase in graviton-photon system}
Similarly to the axion-photon system, gravitons can convert to photons in the presence of an external magnetic field. The interaction between the graviton field and the EM field is described by \cite{Raffelt:1987im}
\begin{equation}
\mathcal{L}_{\mathrm{int}}=-\frac{\kappa}{2}h_{\mu\nu}T^{\mu\nu}, ~~~T_{\mu\nu}= F^{\mu\alpha}F^\nu_{~\alpha}-\frac{1}{4}\eta^{\mu\nu}F_{\alpha\beta}F^{\alpha\beta},
\label{graviton1}
\end{equation}
where $\kappa=\sqrt{16\pi G}=2/M_{\mathrm{Pl}}$, with $G$ the gravitational constant and $M_{\mathrm{Pl}}$, the Planck mass, $h_{\mu\nu}$ denotes the spin-$2$ graviton field, and $T_{\mu\nu}$ is the EM energy-momentum tensor. In the presence of a transverse magnetic field, one photon polarization mixes with a single graviton polarization, reducing the system to an effective two-level problem \cite{Raffelt:1987im}
\begin{equation}
i\frac{d}{dz}\begin{pmatrix}
A\\
h
\end{pmatrix} =\begin{pmatrix}
\Delta_\gamma & \Delta_{g\gamma}\\
\Delta_{g\gamma} & \Delta_g
\end{pmatrix}  \begin{pmatrix}
A\\
h
\end{pmatrix}, 
\label{graviton2}
\end{equation}
where $\Delta_\gamma=-\omega^2_{\mathrm{pl}}/(2\omega)$, $\Delta_g=0$ for massless gravitons in general relativity, and $\Delta_{g\gamma}=\kappa B_T/2 = B_T/M_{\mathrm{Pl}}$.

If the transverse magnetic field rotates adiabatically in the transverse plane, the Hamiltonian acquires a phase in the off-diagonal elements,
\begin{equation}
H=
 \begin{pmatrix}
\Delta_\gamma & \Delta_{g\gamma}e^{-i\varphi}\\
\Delta_{g\gamma}e^{i\varphi} & \Delta_g
 \end{pmatrix},   
 \label{graviton3}
\end{equation}
where $\varphi$ parametrizes the orientation of the magnetic field. The corresponding mixing angle is given by
\begin{equation}
\tan2\theta=\frac{2\Delta_{g\gamma}}{\Delta_\gamma-\Delta_g},    
\end{equation}
hence, the Berry phase becomes 
\begin{equation}
\begin{split}
 \gamma_{\pm}=\mp 2\pi\sin^2\theta\simeq \mp2\pi \Big(\frac{\Delta_{g\gamma}}{\Delta_\gamma-\Delta_g}\Big)^2\simeq \mp 2\pi\Big(\frac{2\omega B_T}{M_{\mathrm{Pl}}\omega^2_{\mathrm{pl}}}\Big)^2\\\mathrm{(mod~~2\pi)},
 \end{split}
 \label{graviton4}
\end{equation}

where in the last two steps we consider the small-mixing limit. Thus, the Berry phase induced by graviton-photon mixing is extremely suppressed by the Planck scale. 

For a representative electron density $n_e\sim 1~\mathrm{cm}^{-3}$, the plasma frequency is $\omega_{\mathrm{pl}}=\sqrt{n_e e^2/m_e}\sim 3.7\times 10^{-11}~\mathrm{eV}$. This leads to an oscillation scale $\Delta_{\mathrm{osc}}\sim 5.7\times 10^{-22}~\mathrm{eV}$, which is significantly smaller than the typical phase variation rate $d\varphi/dz\sim 2\pi/L_{\mathrm{rot}}\sim 10^{-6}~\mathrm{eV}$ for a laboratory setup with $L_{\mathrm{rot}}\sim 1~\mathrm{m}$. Consequently, the adiabaticity condition $d\varphi/dz \ll \Delta_{\mathrm{osc}}$ is strongly violated, rendering the effect unobservable in terrestrial interferometric experiments.

If gravitons acquire a small mass $m_g$, the diagonal term becomes $\Delta_g\simeq -m_g^2/(2\omega)$, and in the small-mixing and ultra-vacuum regime the Berry phase is given by
\begin{equation}
\gamma_{\pm}
\simeq
\mp 2\pi
\left(\frac{2\omega B_T}{M_{\mathrm{Pl}}m_g^2}\right)^2~~\mathrm{(mod~~2\pi)}.
\label{graviton5}
\end{equation}
Although a small graviton mass can formally enhance the Berry phase, this regime is not experimentally accessible due to the breakdown of adiabaticity. For \(m_g\sim 10^{-24}\,\mathrm{eV}\) \cite{Wu:2023rib} and \(\omega\sim 1.2\,\mathrm{eV}\), the oscillation scale is \(\Delta_{\mathrm{osc}}\sim m_g^2/(2\omega)\sim 4\times 10^{-49}\,\mathrm{eV}\), requiring \(d\varphi/dz \ll 4\times 10^{-49}\,\mathrm{eV}\) for adiabatic evolution (see Eq.~\eqref{cm4}). However, in a realistic laboratory setup with magnetic-field variation over \(L_{\mathrm{rot}}\sim 1\,\mathrm{m}\), one has \(d\varphi/dz\sim 2\pi/L_{\mathrm{rot}}\sim 10^{-6}\,\mathrm{eV}\), which violates the adiabaticity condition by many orders of magnitude. Consequently, despite a formally large Berry phase, such effects cannot be probed in laboratory interferometers, because of the violation of the adiabaticity condition.

\section{Axion-induced Berry phase in presence of axion-quasiparticle}\label{sec4}

\begin{figure}[ht]
    \centering
\includegraphics[width=3in]{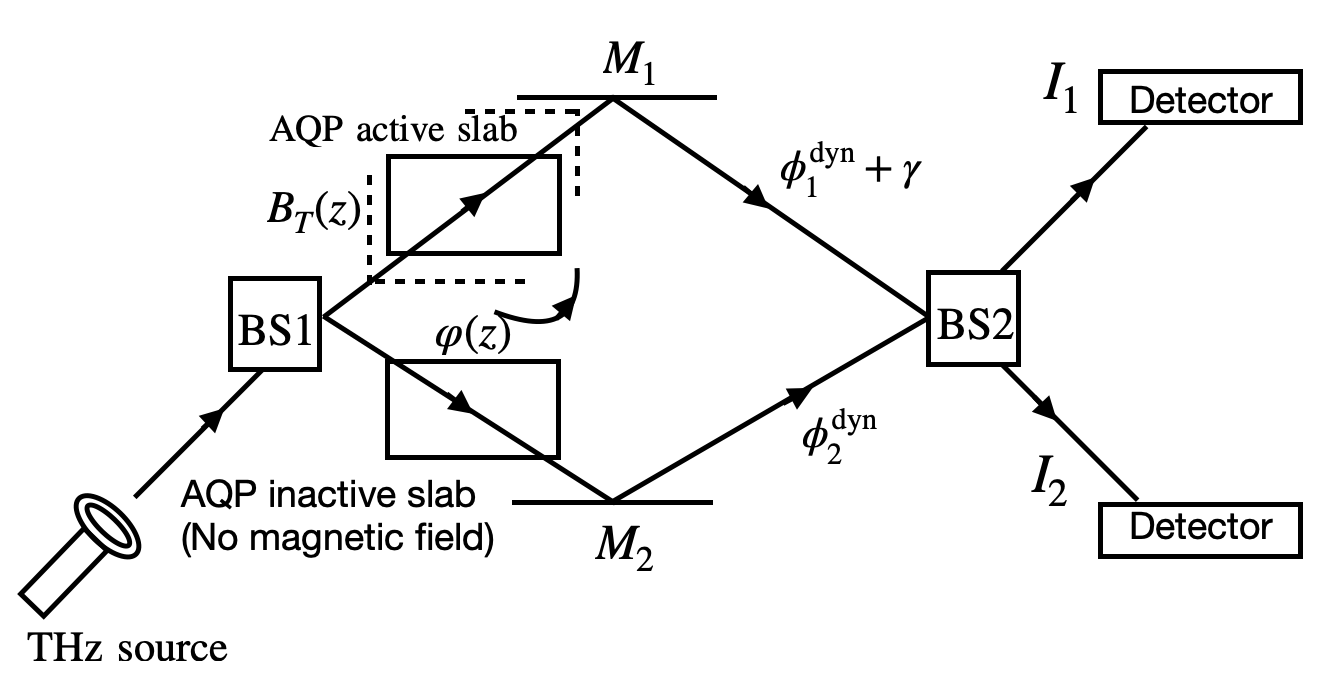}
   \caption{Schematic representation of detecting axion-induced Berry phase in presnece of axion quasiparticle using THz Mach-Zehnder interferometer}
    \label{plot3}
\end{figure}

Axion quasiparticles (AQPs) have been proposed as collective spin-wave longitudinal excitations in antiferromagnetic topological insulators. These modes couple to the EM Chern-Simons term and, in the presence of an external magnetic field, exhibit mass mixing with the electric field. This mechanism forms the basis of a proposed DM detection scheme known as TOORAD (TOpolOgical Resonant Axion Detection) \cite{SchutteEngel:2021}. In Ref.~\cite{Li2010}, it was first suggested that AQPs could be realized in iron-doped bismuth selenide, $(\mathrm{Bi}_{1-x} \mathrm{Fe}_x)_2 \mathrm{Se}_3$. More recently, the compound 2D $\mathrm{Mn}\mathrm{Bi}_2\mathrm{Te}_4$ has emerged as a promising candidate of AQP topological insulator \cite{Qiu:2025bbi}.

In such systems, the axion quasiparticle can hybridize with the electric field, giving rise to axion-polariton modes \cite{Li2010}. The polariton frequency is determined by the magnetic anisotropy scale of the antiferromagnet, typically of order $\mathcal{O}(\mathrm{meV})$ corresponding to the low-THz frequency range. Importantly, this frequency can be tuned by applying an external static magnetic field $B$. Consequently, antiferromagnetic topological insulators hosting AQPs provide a promising platform for probing axion DM in the $\mathrm{meV}$ mass range, a regime that is challenging to access with conventional resonant detection techniques such as THz cavities, because of the lack of large volume, low noise, and large bandwidth \cite{SchutteEngel:2021}.

A schematic representation of the measurement of an axion-induced Berry phase in the presence of an AQP using a MZI is shown in FIG.~\ref{plot3}. The overall configuration is analogous to the axion-photon case, with the key difference that the optical source operates in the THz regime and the interferometer is correspondingly adapted for THz frequencies. The incident THz beam (millimeter wave) is split into two paths by the BS1. In one arm, the beam traverses a topological material hosting AQPs, subjected to a transverse magnetic field. The magnetic field is varied adiabatically along the propagation direction to ensure the accumulation of a Berry phase. The second arm serves as a reference and contains an AQP-inactive slab without any applied magnetic field, thereby isolating material-induced dynamical effects while avoiding additional geometric phase contributions. After reflection from mirrors $M_1$ and $M_2$, the two beams are recombined at the BS2, and the resulting interference pattern is recorded at the detector. The measured signal contains both dynamical and geometric phase contributions. A similar experiment performed in the absence of the magnetic field yields only the dynamical phase difference. The Berry phase can then be extracted from the differential fringe shift between these two configurations. 

The AQP, denoted by $\delta\Theta$, can be interpreted as a dynamical fluctuation of the axion angle $\Theta$. In topological insulators, the EM response contains the topological $\Theta$-term in the effective action, given by \cite{Li2010}
\begin{equation}
S_\theta=\frac{\alpha}{\pi}\int d^4x\, \Theta\,\mathbf{E}\cdot \mathbf{B},
\label{aqp1}
\end{equation}
where $\alpha$ is the fine-structure constant. This term describes the topological magnetoelectric response of the system.

For ordinary time-reversal invariant topological insulators, the axion angle $\Theta$ takes quantized values $0$ or $\pi$, determined by the topology of the electronic band structure. In contrast, in antiferromagnetic topological insulators, the magnetic order breaks time-reversal symmetry and allows the axion angle to become dynamical. In this case the axion angle can be written as
\begin{equation}
\Theta(x,t)=\Theta^0+\delta\Theta(x,t),
\label{aqp2}
\end{equation}
where $\Theta^0$ represents the static background value and $\delta\Theta(x,t)$ corresponds to the dynamical fluctuation identified as the AQP.

Substituting Eq.~\eqref{aqp2} into Eq.~\eqref{aqp1}, the topological action becomes \cite{SchutteEngel:2021}
\begin{equation}
S_{\theta}=\frac{\alpha}{\pi}\int d^4x\, (\Theta^0+\delta\Theta)\mathbf{E}\cdot \mathbf{B}.
\label{aqp3}
\end{equation}

The background term $\Theta^0\,\mathbf{E}\cdot\mathbf{B}$ describes the static topological magnetoelectric effect, which gives rise to phenomena such as the surface Hall response and the quantized magnetoelectric polarization. The fluctuation term $\delta\Theta\,\mathbf{E}\cdot\mathbf{B}$ represents the interaction between the AQP and the EM field.

In antiferromagnetic topological insulators, the AQP originates from fluctuations of the antiferromagnetic order parameter. The magnetic order is characterized by the Neel vector $\mathbf{L}=\mathbf{M}_1-\mathbf{M}_2$,
where $\mathbf{M}_{1}$ and $\mathbf{M}_{2}$ denote the magnetizations of the two magnetic sublattices. In the antiferromagnetic phase $\mathbf{L}\neq0$, whereas $\mathbf{L}=0$ in the paramagnetic phase. Dynamical longitudinal fluctuations of the Neel vector modulate the axion angle $\Theta$, therefore, giving rise to the AQP excitation.

The AQP is described by an effective action that includes its interaction with the EM field as \cite{Li2010, Sekine:2014xva}
\begin{equation}
\begin{split}
S_{\delta\Theta}&=\frac{f^2_{\Theta}}{2}\int dt d^3r\Big[(\partial_\mu \delta \Theta)(\partial^\mu\delta\Theta)-m^2_{\Theta}\delta\Theta^2
\Big]+\\
&\frac{\alpha}{\pi}\int dtd^3r\,\delta\Theta \mathbf{E}\cdot\mathbf{B},
\end{split}
\end{equation}
where $m_\Theta$ denotes the mass of the AQP and $f_\Theta$ represents the stiffness parameter, which plays the role of an effective decay constant controlling the strength of the AQP-photon interaction. 

The coupling between the AQP and the EM field allows the AQP to decay into two photons. The corresponding decay width is given by \cite{SchutteEngel:2021}
\begin{equation}
\Gamma_{\delta\Theta\gamma\gamma}=
\frac{\alpha^2}{256\pi^3}\frac{m_\Theta^3}{f_\Theta^2}
\simeq
6.7\times10^{-22}\,\mathrm{eV}
\left(\frac{m_\Theta}{\mathrm{1~meV}}\right)^3
\left(\frac{100\,\mathrm{eV}}{f_\Theta}\right)^2.
\label{aqp4}
\end{equation}
The corresponding lifetime is on the order of weeks for the benchmark parameters shown above. This timescale is much longer than the characteristic timescale ($\sim \mathcal{O}$ (min)) relevant for the processes considered in the present analysis, and therefore the decay of the AQP can be safely neglected. The AQP parameters $m_\Theta$ and $f_\Theta$ can be experimentally determined for specific materials. These quantities can be inferred from measurements of the axion-polariton resonance and the associated energy gap using transmission spectroscopy.

The mixing between axions, AQPs, and photons modifies the Maxwell equations through the couplings of the axion and AQP fields to the EM field. The resulting equations of motion consist of the modified Maxwell equations together with the Klein-Gordon equations for the AQP and axion fields, which can be written as \cite{Li2010,SchutteEngel:2021}
\begin{equation}
\begin{split}
\nabla\cdot \mathbf{D} = \rho_f - \frac{\alpha}{\pi}\nabla(\delta\Theta+\Theta^0)\cdot \mathbf{B} - g_{a\gamma\gamma}\nabla a\cdot \mathbf{B}, \\
\nabla\times \mathbf{H}-\partial_t \mathbf{D} = \mathbf{J}_f
+\\
\frac{\alpha}{\pi}\left[\mathbf{B}\partial_t(\delta\Theta+\Theta^0)-
\mathbf{E}\times\nabla(\delta\Theta+\Theta^0)\right]+\\
g_{a\gamma\gamma}\left(\mathbf{B}\partial_t a-\mathbf{E}\times\nabla a\right), \\
\nabla\cdot \mathbf{B} =0, \\
\nabla\times \mathbf{E} = -\partial_t \mathbf{B}, \\
\partial_t^2\delta\Theta - \sum_i v_i^2\partial_i^2\delta\Theta + m_\Theta^2 \delta\Theta = \Lambda\mathbf{E}\cdot \mathbf{B}, \\
(\partial_t^2-\nabla^2+m_a^2)a = g_{a\gamma\gamma}\mathbf{E}\cdot \mathbf{B},
\end{split}
\end{equation}
where $\Lambda=\alpha/\pi f^2_\Theta$, $v_i$ denotes the spin-wave velocity, $\mathbf{D}$ is the electric displacement field, and $\mathbf{H}$ is the magnetic field strength. The parameters $\epsilon$ and $\mu$ denote the scalar permittivity and permeability of the medium, respectively, while $\rho_f$ and $\mathbf{J}_f$ denote the free charge and current densities.

We now assume the presence of a strong static external magnetic field $\mathbf{B}_e$, oriented along the $y$-direction, $\mathbf{B}_e = B_e \hat{y}$. Linearizing the above equations around this background field and keeping only the leading order terms yields the following one dimensional equations of motion \cite{SchutteEngel:2021}

\begin{equation}
\begin{aligned}
(\partial_z^2-n^2\partial_t^2-\sigma\mu\partial_t)E_y
&=\mu B_e \partial_t^2\left(\frac{\alpha}{\pi}\delta\Theta+g_{a\gamma\gamma}a\right),\\
(v_z^2\partial_z^2-\partial_t^2-m_\Theta^2)\delta\Theta
&=-\Lambda B_e E_y,\\
(\partial_z^2-\partial_t^2-m_a^2)a
&=-g_{a\gamma\gamma}B_eE_y,
\end{aligned}
\end{equation}

where $n=\sqrt{\mu\epsilon}$ is the refractive index of the medium, $\sigma$ is the electrical conductivity of the material, and $v_z$ denotes the spin-wave velocity along the $z$-direction. Furthermore, we assume that no free surface charges or currents are present at the interfaces.

Upon Fourier transforming along the propagation direction, $\partial_z^2\rightarrow -k^2$, the axion-AQP-photon system forms a coupled three-level system. The linearized dynamics of the relevant fields can be expressed in terms of the state vector $\mathbf{X} \equiv (E_y~~\delta\Theta~~a)^T$ .

Including dissipation effects, the equations of motion can be written in compact matrix form as

\begin{equation}
K\partial_t^2 \mathbf{X} + \Gamma\partial_t \mathbf{X} + M\mathbf{X} = 0 ,
\label{matrix1}
\end{equation}

where $K$, $\Gamma$, and $M$ are $3\times3$ matrices given by

\begin{equation}
K=
\begin{pmatrix}
1 & \dfrac{\alpha B_e}{\pi\epsilon} & \dfrac{g_{a\gamma\gamma}B_e}{\epsilon} \\
0 & 1 & 0 \\
0 & 0 & 1
\end{pmatrix},
\qquad
\Gamma=
\begin{pmatrix}
\Gamma_p & \Gamma_{\times,1} & 0 \\
\Gamma_{\times,2} & \Gamma_m & 0 \\
0 & 0 & 0
\end{pmatrix},
\label{matrix2}
\end{equation}

\begin{equation}
M=
\begin{pmatrix}
\dfrac{k^2}{n^2} & 0 & 0 \\
-\Lambda B_e & v_z^2 k^2 + m_\Theta^2 & 0 \\
-g_{a\gamma\gamma} B_e & 0 & k^2 + m_a^2
  \end{pmatrix}.
\label{matrix3}
\end{equation}

Here $\Gamma_p=\sigma/\epsilon$ represents the photon loss rate, $\Gamma_m$ denotes the equivalent magnon loss rate, and $\Gamma_{\times,1}$ and $\Gamma_{\times,2}$ correspond to mixed losses arising from photon-magnon interactions. No loss term is included for the axion field, since the axion is assumed to constitute DM and is effectively stable on experimental timescales.

The corresponding wave number $k$ is determined from the dispersion relation of the system, given by
\begin{equation}
\left(k^2 - \frac{k_a^2 + k_\Theta^2}{2}\right)^2
= b_a^2 k_p^2 + \left(\frac{k_a^2 - k_\Theta^2}{2}\right)^2,
\label{disp1}
\end{equation}

where
\begin{equation}
\begin{split}
k_\Theta^2 = n^2 \omega^2 \left(1 - \frac{b^2}{\omega^2 - m_\Theta^2}\right), \quad
b^2 = \frac{\alpha}{\pi} \frac{\Lambda B_e^2}{\epsilon}, \\
b_a^2 = \frac{g_{a\gamma\gamma}^2 B_e^2}{\epsilon}, \quad
k_p^2 = n^2 \omega^2.
\end{split}
\label{disp2}
\end{equation}

Although the field content is three-component, $\mathbf{X}=(E_y~~\delta\Theta~~a)^T$, the dispersion relation in Eq.~\eqref{disp1} is already written in the reduced TOORAD limit $v_z\simeq 0$. In this limit, the AQP equation loses its $k$-dependence and $\delta\Theta$ no longer represents an independent propagating mode along the $z$-direction; instead, it is algebraically determined by the photon field. Consequently, the spatial propagation problem reduces to two branches, $k_\pm$, rather than three independent propagating wave numbers. For finite spin-wave velocity $v_z\neq 0$, one would in general recover three modes.

To leading order in the axion-photon coupling, the dispersion relation yields
\begin{equation}
k_+ = k_\Theta + \mathcal{O}(g_{a\gamma\gamma}^2), \qquad
k_- = k_a + \mathcal{O}(g_{a\gamma\gamma}^2).
\label{soln}
\end{equation}

To calculate the Berry phase in the TOORAD system, we consider the lossless scenario for simplicity. Using $X(t)=ue^{-i\omega t}$ in Eq.~\eqref{matrix1}, we obtain 
\begin{equation}
(M-\omega^2 K)u=0. 
\label{ck1}
\end{equation}
We can write the mode equation by rotating the external magnetic field adiabatically along the transverse direction as 
\begin{equation}
A(\varphi)u=0, ~~~A(\varphi)=M(\varphi)-\omega^2 K(\varphi)
\label{ck2}
\end{equation}
with
\begin{equation}A(\varphi)=
\begin{pmatrix}
\Delta_\gamma  & -\omega^2\frac{\alpha B_e}{\pi\epsilon}e^{-i\varphi} & -\omega^2\frac{g_{a\gamma\gamma}B_e}{\epsilon}e^{-i\varphi}\\
-\Lambda B_e e^{i\varphi} & \Delta_m & 0\\
-g_{a\gamma\gamma}B_e e^{i\varphi} & 0 & \Delta_a 
\end{pmatrix},    
\label{ck3}
\end{equation}
where 
\begin{equation}
\Delta_\gamma=\frac{k^2}{n^2}-\omega^2,~\Delta_m=v^2_zk^2+m^2_{\Theta}-\omega^2,~\Delta_a=k^2_a+m^2_a-\omega^2.  
\label{ck4}
\end{equation}

For the geometric phase, however, we consider the adiabatic transport of the physical propagating eigenmodes in a canonically normalized basis. In the lossless limit, this may be described by an effective Hermitian propagation Hamiltonian.
Equivalently, one may view this as a constant diagonal rescaling of the algebraic mode variables, $u=S\psi$, chosen to render the mode coupling matrix Hermitian. Since $S$ is $\varphi$-independent, the rescaling itself does not generate any additional geometric term. If one works in the original variables $u$, the corresponding inner product is $\langle u_1|\eta|u_2\rangle$ with $\eta=(S^{-1})^\dagger S^{-1}$; working directly in the canonically normalized basis in which $H(\varphi)$ acts is therefore equivalent and more convenient. Therefore, we obtain

\begin{equation}
H(\varphi)=S^{-1}A(\varphi)S=\begin{pmatrix}
\Delta_\gamma & -G_m e^{-i\varphi} & -G_a e^{-i\varphi}\\
-G_m e^{i\varphi} & \Delta_m & 0\\
-G_a e^{i\varphi} & 0 & \Delta_a
\end{pmatrix},  
\label{ck9}
\end{equation}
where 
\begin{equation}
\begin{split}
G_m=\omega^2\frac{\alpha B_e}{\pi\epsilon}\Big(\frac{\Lambda\pi\epsilon}{\omega^2\alpha}\Big)^{1/2}=\omega B_e\sqrt{\frac{\alpha\Lambda}{\pi\epsilon}},\\    
G_a=g_{a\gamma\gamma}B_e\Big(\frac{\omega^2}{\epsilon}\Big)^{1/2}=\frac{\omega g_{a\gamma\gamma}B_e}{\sqrt{\epsilon}},\\
\end{split}
\label{ck10}
\end{equation}
and the diagonal matrix becomes
\begin{equation}
S=\mathrm{diag}\Big(1, \sqrt{\frac{\Lambda\pi\epsilon}{\omega^2\alpha}}, \sqrt{\frac{\epsilon}{\omega^2}}\Big). 
\label{ck11}
\end{equation}

For fixed $\omega$ and local field orientation $\varphi$, the lossless normal modes are determined by the kernel condition $A(\varphi)u=0$. After the rescaling $u=S\psi$, this becomes $H(\varphi)\psi=0$, so $A(\varphi)$ and $H(\varphi)$ encode the same instantaneous mode composition and are related by a $\varphi$-independent similarity transformation. In a local-mode (WKB) description, when the field direction varies sufficiently slowly along the propagation direction, $\varphi=\varphi(z)$, a wave packet launched in one branch remains in the corresponding instantaneous eigenvector up to an overall phase. It is therefore convenient to determine the adiabatically transported branches from the Hermitian spectral problem of $H(\varphi)$, with the physically relevant branch selected by continuity with the corresponding solution of the local dispersion relation. The geometric phase is then computed from these instantaneous eigenstates rather than from generic eigenvectors of the algebraic mode matrix in Eq.~\eqref{ck3}.

Thus, the instantaneous physical eigenstates satisfy
\begin{equation}
H(\varphi)|n(\varphi)\rangle =\lambda_n|n(\varphi)\rangle,~~~|n(\varphi)\rangle=\begin{pmatrix}
 u_\gamma(\varphi)\\
 u_m(\varphi)\\
 u_a(\varphi)
\end{pmatrix} ,   
\label{ck12}
\end{equation}
where $|n(\varphi)\rangle$ is the eigenvector corresponding to the instantaneous eigenvalue $\lambda_n$. Therefore, the eigenvalue equations become
\begin{equation}
\begin{split}
(\Delta_\gamma-\lambda_n)u_\gamma(\varphi)-G_me^{-i\varphi}u_m(\varphi)-G_ae^{-i\varphi}u_a(\varphi)=0,\\
-G_me^{i\varphi}u_\gamma(\varphi)+(\Delta_m-\lambda_n)u_m(\varphi)=0,\\
-G_ae^{i\varphi}u_\gamma(\varphi)+(\Delta_a-\lambda_n)u_a(\varphi)=0.
\end{split}
\label{ck13}
\end{equation}
We express $u_m(\varphi)$ and $u_a(\varphi)$ in terms of $u_\gamma(\varphi)$ using  last two equations of Eq.~\eqref{ck13} as
\begin{equation}
u_m(\varphi)=\frac{G_me^{i\varphi}}{(\Delta_m-\lambda_n)}u_\gamma(\varphi), ~~~  u_a(\varphi)=\frac{G_ae^{i\varphi}}{(\Delta_a-\lambda_n)}u_\gamma(\varphi), 
\label{ck14}
\end{equation}
and the eigenvector can be written as
\begin{equation}
|n(\varphi)\rangle=u_\gamma(\varphi)\begin{pmatrix}
1\\
\frac{G_me^{i\varphi}}{(\Delta_m-\lambda_n)}\\
\frac{G_ae^{i\varphi}}{(\Delta_a-\lambda_n)}
\end{pmatrix}.  
\label{ck15}
\end{equation}
We obtain the expression for $u_\gamma(\varphi)$ using the normalization condition of $|n(\phi)\rangle$, i.e., $\langle n(\varphi)|n(\phi)\rangle=1$ from Eq.~\eqref{ck15} as
\begin{equation}
u_\gamma(\varphi)=\frac{1}{\sqrt{1+\frac{G^2_m}{(\Delta_m-\lambda_n)^2}+\frac{G^2_a}{(\Delta_a-\lambda_n)^2}}},   
\label{ck16}
\end{equation}
up to an overall irrelevant phase choice. Therefore, the normalized eigenvector becomes
\begin{equation}
|n(\varphi)\rangle=  \frac{1}{\sqrt{1+\frac{G^2_m}{(\Delta_m-\lambda_n)^2}+\frac{G^2_a}{(\Delta_a-\lambda_n)^2}}}\begin{pmatrix}
1\\
\frac{G_me^{i\varphi}}{(\Delta_m-\lambda_n)}\\
\frac{G_ae^{i\varphi}}{(\Delta_a-\lambda_n)}
\end{pmatrix}.  
\label{ck17}
\end{equation}
To obtain the Berry phase, we write the Berry connection for the $n$'th mode as
\begin{equation}
\mathcal{A}_n(\varphi)=i\langle n(\varphi)|\partial_\varphi n(\varphi)\rangle .
\label{ck18}
\end{equation}
Using Eq.~\eqref{ck17} we obtain 
\begin{equation}
\mathcal{A}_n(\varphi)=-\frac{\frac{G^2_m}{(\Delta_m-\lambda_n)^2}+\frac{G^2_a}{(\Delta_a-\lambda_n)^2}}{1+\frac{G^2_m}{(\Delta_m-\lambda_n)^2}+\frac{G^2_a}{(\Delta_a-\lambda_n)^2}}.
\label{ck19}
\end{equation}
Since, the right hand side of Eq.~\eqref{ck19} is independent of $\varphi$, the Berry phase for one cycle ($\varphi: 0\to 2\pi$) becomes
\begin{equation}
\gamma_n=\int ^{2\pi}_0 \mathcal{A}_n(\varphi)d\varphi=-2\pi \frac{\frac{G^2_m}{(\Delta_m-\lambda_n)^2}+\frac{G^2_a}{(\Delta_a-\lambda_n)^2}}{1+\frac{G^2_m}{(\Delta_m-\lambda_n)^2}+\frac{G^2_a}{(\Delta_a-\lambda_n)^2}},
\label{ck20}
\end{equation}
which can be written equivalently as
\begin{equation}
\gamma_n\equiv 2\pi\Bigg[1+\frac{G^2_m}{(\Delta_m-\lambda_n)^2}+\frac{G^2_a}{(\Delta_a-\lambda_n)^2}\Bigg]^{-1}, ~~~\mathrm{(mod~~2\pi)}.
\label{ck21}
\end{equation}

To obtain the eigenvalues $\lambda_n$, we use Eq.~\eqref{ck14} and write the first equation of Eq.~\eqref{ck13} as
\begin{equation}
\begin{split}
(\Delta_\gamma-\lambda_n)(\Delta_m-\lambda_n)(\Delta_a-\lambda_n)-G^2_m(\Delta_a-\lambda_n)-\\G^2_a(\Delta_m-\lambda_n)=0.  
\end{split}
\label{ck22}
\end{equation}
This is a cubic equation in $\lambda_n$ and can be solved analytically. However, the interesting scenario is to find the eigenvalues and hence the Berry phase in five different limiting cases. 

(i) $G_a=0:$ in this limit, the eigenvalues are obtained as
\begin{equation}
\begin{split}
\lambda_1&=\Delta_a,\\
\lambda_{2,3}&=\frac{\Delta_m+\Delta_\gamma}{2}\pm \frac{1}{2}\sqrt{(\Delta_m-\Delta_\gamma)^2+4G^2_m}. 
\end{split}
\label{ck23}
\end{equation}
This limit reduces the TOORAD system in an effective photon-AQP two level scenario, where the axion is effectively decoupled ($g_{a\gamma\gamma}\to 0$) from the system. 

Therefore, for the decoupled axion state $(\lambda_1=\Delta_a)$, it does not acquire a geometric phase and hence
\begin{equation}
 \gamma_1=0,~~~\mathrm{(mod~2\pi)}.
\end{equation}
Similarly for the eigenvalues $\lambda_{2,3}$ we obtain the Berry phases
\begin{equation}
\begin{split}
\gamma_2=\pi\Bigg(1-\frac{\Delta_m-\Delta_\gamma}{\sqrt{(\Delta_m-\Delta_\gamma)^2+4G_m^2}}\Bigg),~~~\mathrm{(mod~2\pi)}\\
\gamma_3=\pi\Bigg(1+\frac{\Delta_m-\Delta_\gamma}{\sqrt{(\Delta_m-\Delta_\gamma)^2+4G^2_m}}\Bigg)~~~\mathrm{(mod~2\pi)}.
\end{split}
\end{equation}

(ii) $G_m=0:$ In this limit, the eigenvalues are obtained as
\begin{equation}
\begin{split}
\lambda_1&=\Delta_m,\\
\lambda_{2,3}&=\frac{\Delta_a+\Delta_\gamma}{2}\pm \frac{1}{2}\sqrt{(\Delta_a-\Delta_\gamma)^2+4 G^2_a}.
\end{split}
\label{ck24}
\end{equation}
This limit reduces the TOORAD system in an effective photon-axion two level scenario, where the AQP is effectively decoupled $(\Lambda\to 0)$ from the system. 

In this case, the AQP decoupled state $(\lambda_1=\Delta_m)$ does not acquire a Berry phase and hence
\begin{equation}
\gamma_1=0~~\mathrm{(mod~2\pi)},    
\end{equation}
and for $\lambda_{2,3}$, the Berry phases become
\begin{equation}
\begin{split}
\gamma_2=\pi\Bigg(1-\frac{\Delta_a-\Delta_\gamma}{\sqrt{(\Delta_a-\Delta_\gamma)^2+4G_a^2}}\Bigg),~~\mathrm{(mod~2\pi)}\\
\gamma_3=\pi\Bigg(1+\frac{\Delta_a-\Delta_\gamma}{\sqrt{(\Delta_a-\Delta_\gamma)^2+4G^2_a}}\Bigg)~~~\mathrm{(mod~2\pi)}.
\end{split}    
\end{equation}

(iii) $\Delta_m=\Delta_a=\Delta_s:$ In this limit, the AQP and axions are energetically degenerate. Therefore, the eigenvalues become
\begin{equation}
\begin{split}
\lambda_1 &=\Delta_s,\\
\lambda_{2,3}&= \frac{\Delta_s+\Delta_\gamma}{2}\pm \frac{1}{2}\sqrt{(\Delta_s-\Delta_\gamma)^2+4(G^2_m+G^2_a)}.
\end{split}   
\label{ck25}
\end{equation}

The dark state corresponds to $\lambda_1=\Delta_s$ is orthogonal to the photon-coupled bright combination and hence for the dark state
\begin{equation}
\gamma_1=0~~~\mathrm{(mod~~2\pi)}.  
\end{equation}
Similarly, for the other two eigenvalues $\lambda_{2,3}$, the corresponding Berry phases become
\begin{equation}
\begin{split}
\gamma_2=\pi\Bigg(1-\frac{\Delta_s-\Delta_\gamma}{\sqrt{(\Delta_s-\Delta_\gamma)^2+4(G^2_m+G^2_a)}}\Bigg),~\mathrm{(mod~~2\pi)}\\
\gamma_3=\pi\Bigg(1+\frac{\Delta_s-\Delta_\gamma}{\sqrt{(\Delta_s-\Delta_\gamma)^2+4(G^2_m+G^2_a)}}\Bigg)~~\mathrm{(mod~~2\pi)}.
\end{split}
\end{equation}

(iv) $\frac{G_m}{\Delta_m-\Delta_\gamma}\ll 1, \frac{G_a}{\Delta_a-\Delta_\gamma}\ll1 :$ This limit corresponds to the small mixing case, which results the eigenvalues as
\begin{equation}
\begin{split}
\lambda_1&\approx\Delta_\gamma-\frac{G^2_a}{\Delta_a-\Delta_\gamma}-\frac{G^2_m}{\Delta_m-\Delta_\gamma},\\
\lambda_2&\approx\Delta_m+\frac{G^2_m}{\Delta_m-\Delta_\gamma},\\
\lambda_3&\approx \Delta_a+\frac{G^2_a}{\Delta_a-\Delta_\gamma}.
\end{split}
\label{ck26}
\end{equation}

For a three-level system, there exist three instantaneous eigenstates, each associated with a distinct eigenvalue and corresponding Berry phase. Among these, we focus on the eigenmode that is predominantly photon-like, as it is the only mode directly accessible in the experimental setup. In this regime, the eigenstate propagates primarily as a photon, with only small admixtures of the AQP and axion components. In the small-mixing limit, the corresponding eigenvalue approaches $\lambda_n\approx \Delta_\gamma$ from Eq.~\eqref{ck26}. Consequently, the magnitude of the associated Berry phase, as given in Eq.~\eqref{ck21}, can be evaluated in this limit as 
\begin{equation}
\gamma^{(\gamma)}\simeq -2\pi\Bigg[\frac{G^2_m}{(\Delta_m-\Delta_\gamma)^2}+\frac{G^2_a}{(\Delta_a-\Delta_\gamma)^2}\Bigg], ~~~\mathrm{(mod~2\pi)}  
\label{ck27}
\end{equation}
To obtain the Berry phases for the AQP and the axion mode, we have to take the corresponding AQP and axion component as the reference component for each cases i.e., (21) entry and (31) entry for the two scenarios would be normalized with one in Eq.~\eqref{ck15}, and similarly obtain the Berry phase in the small mixing limit for AQP and axion modes as
\begin{equation}
\begin{split}
\gamma^{(m)}&\simeq 2\pi \frac{G^2_m}{(\Delta_m-\Delta_\gamma)^2},~~~\mathrm{(mod~2\pi)}\\
\gamma^{(a)}&\simeq 2\pi \frac{G^2_a}{(\Delta_a-\Delta_\gamma)^2}~~~~~\mathrm{(mod~2\pi)}.
\end{split}
\label{ck28}
\end{equation}
The difference in sign between the Berry phases in Eqs.~\eqref{ck27} and \eqref{ck28} arises from the redistribution of the photon component among the eigenstates. In particular, the photon-like state predominantly loses weight to the other sectors, while the AQP and axion states acquire small photon admixtures. This asymmetry in the composition of the eigenstates leads to opposite signs in the corresponding Berry phases.

(v) $\Delta_\gamma\simeq \Delta_m\simeq \Delta_a:$ In this regime, it is convenient to introduce the orthonormal bright $(|b\rangle)$ and dark $(|d\rangle)$ states, which are combinations of the AQP state $(|m\rangle)$ and the axion state $(|a\rangle)$, given as
\begin{equation}
|b\rangle=\frac{G_m|m\rangle+G_a|a\rangle}{G},~~~|d\rangle=\frac{-G_a|m\rangle+G_m|a\rangle}{G},    
\end{equation}
where $G=\sqrt{G^2_m+G^2_a}$. 

In the basis $(|\gamma\rangle, |b\rangle, |d\rangle)$, the corresponding unitary transformation is
\begin{equation}
U=
\begin{pmatrix}
1 &0 &0\\
0& G_m/G & G_a/G\\
0 &-G_a/G & G_m/G
\end{pmatrix}.  
\end{equation}
The Hamiltonian in the bright-dark basis is then given by
\begin{equation}
H^\prime(\varphi)=U   H(\varphi) U^\dagger=
\begin{pmatrix}
\Delta_\gamma & G e^{-i\varphi} & 0\\
Ge^{i\phi} & \Delta_b & \delta_{bd}\\
0 &\delta_{bd} & \Delta_d
\end{pmatrix},
\end{equation}
where
\begin{equation}
\begin{split}
\Delta_b=\frac{G_m^2\Delta_m+G_a^2\Delta_a}{G^2_m+G^2_a},~~\Delta_d=\frac{G^2_a\Delta_m+G^2_m\Delta_a}{G^2_m+G^2_a},\\
\delta_{bd}=\frac{G_m G_a}{G^2}(\Delta_a-\Delta_m).~~~~~~~~~~~~~~~~~
\end{split}
\end{equation}
Therefore, in the vicinity of the triple-resonance point, where $(\Delta_\gamma\simeq \Delta_m\simeq \Delta_a)$, the off-diagonal bright-dark mixing $\delta_{bd}$ is parametrically suppressed. As a result, the dark state approximately decouples at leading order, and the dynamics reduces to an effective $2\times 2$ system in the bright sector,
\begin{equation}
H^\prime_{\mathrm{eff}}    (\varphi)=\begin{pmatrix}
\Delta_\gamma & Ge^{-i\varphi}\\
G e^{i\varphi} & \Delta_b
\end{pmatrix}.
\end{equation}

We obtain two bright sector eigenvalues as
\begin{equation}
\lambda_{\pm}=\frac{\Delta_\gamma+\Delta_b}{2}\pm \frac{1}{2}\sqrt{(\Delta_\gamma-\Delta_b)^2+4(G_m^2+G^2_a)},    
\end{equation}
and $\lambda_d\simeq \Delta_d$. The corresponding Berry phases can be obtained as
\begin{equation}
\begin{split}
\gamma_{\pm}=\mp \pi\Bigg[1-\frac{\Delta_\gamma-\frac{G^2_m\Delta_m+G^2_a\Delta_a}{G^2_m+G^2_a}}{\sqrt{\Big(\Delta_\gamma-\frac{G^2_m\Delta_m+G^2_a\Delta_a}{G^2_m+G^2_a}\Big)^2+4(G^2_m+G^2_a)}}\Bigg]\\~~\mathrm{(mod~~2\pi)},  
\end{split}
\label{aqpmain}
\end{equation}
and $\gamma_d\simeq 0$ at the leading order as the dark state is approximately $\varphi$ independent.

For the THz source, we consider a laser with power $P \simeq 10^{-5}~\mathrm{W}$ and frequency $f \simeq 1~\mathrm{THz}$, i.e., $\omega=2\pi f\simeq 4.13~\mathrm{meV}$. 
We further assume an integration time $T \simeq 4~\mathrm{min}$, negligible longitudinal velocity $v_z \simeq 0$,
and $k_a \simeq 0$, appropriate for axion DM. For non-relativistic axion DM, one has $k_a\sim m_a v$; throughout this benchmark we use the spatially homogeneous source approximation $k_aL\ll 1$, so $k_a$ is neglected at leading order. Therefore, the total number of detected photons is $N=PT/\omega\simeq 3.6\times 10^{18}$, which corresponds to the quantum shot-noise limit $\delta\phi_{\mathrm{shot}}\simeq 5.3\times 10^{-10}$.

For benchmark parameters appropriate to a topological magnetic insulator hosting AQPs, and within the regime probed by THz spectroscopy, the relevant material quantities with $m_\Theta\simeq 1.8~\mathrm{meV}$, $f_\Theta\simeq 70~\mathrm{eV}$, $\epsilon\simeq 25$, $\mu\simeq 1$ in presence of $B_e\simeq 2~\mathrm{T}$ external magnetic field can be expressed as
\begin{equation}
\begin{split}
b^2\simeq 6.71\times 10^{-6}~\mathrm{eV^2}\Big(\frac{25}{\epsilon}\Big)\Big(\frac{B_e}{2~\mathrm{T}}\Big)^2\Big(\frac{70~\mathrm{eV}}{f_\Theta}\Big)^2,\\
k^2_\Theta\simeq 4.26\times 10^{-4}~\mathrm{eV^2}\Big(\frac{n^2}{25}\Big)\Big(\frac{\omega}{4.13~\mathrm{meV}}\Big)^2\times\\\Big[1-\frac{b^2/(6.71\times10^{-6})~\mathrm{eV^2}}{(\omega/4.13~\mathrm{meV})^2-(m_\Theta/1.8~\mathrm{meV})^2}\Big].~~~~
\end{split}
\end{equation}

We estimate the values for $G_m$ and $G_a$ for probing axion at the mass $\mathcal{O}(\mathrm{meV})$ as
\begin{equation}
\begin{split}
G_m\simeq 1.06\times 10^{-5}~\mathrm{eV^2}\Big(\frac{25}{\epsilon}\Big)^{\frac{1}{2}} \Big(\frac{\omega}{4.13~\mathrm{meV}}\Big)\Big(\frac{B_e}{2~\mathrm{T}}\Big)\times\\\Big(\frac{70~\mathrm{eV}}{f_\Theta}\Big),\\
G_a\simeq 3.21\times 10^{-22}~\mathrm{eV^2} \Big(\frac{25}{\epsilon}\Big)^{\frac{1}{2}} \Big(\frac{\omega}{4.13~\mathrm{meV}}\Big)\times\\\Big(\frac{B_e}{2~\mathrm{T}}\Big)\Big(\frac{g_{a\gamma\gamma}}{10^{-12}~\mathrm{GeV^{-1}}}\Big),
\end{split}
\end{equation}
and the absolute values of the resonance parameters become
\begin{equation}
\begin{split}
|\Delta_m|\simeq 1.38\times 10^{-5}~\mathrm{eV^2},~~|\Delta_\gamma|\simeq 1.70\times 10^{-5}~\mathrm{eV^2},\\ ~~|\Delta_a|\simeq 9.04\times 10^{-6}~\mathrm{eV^2}, ~~~~~~~~~~~~~~~~~~~~~~
\end{split}
\end{equation}
for $m_a\simeq 2.83~\mathrm{meV}$ and $k\simeq k_a\approx 0$, as we consider the axion is a non-relativistic time-oscillating DM. The above parameter values are obtained for the $k\simeq k_a$ branch.

For the benchmark parameters chosen above, Eq.~\eqref{aqpmain} gives
\begin{equation}
\Delta\gamma \equiv \pi-|\gamma_\pm| \simeq 0.15\pi ,
\end{equation}
indicating a geometric phase that is, in principle, accessible to present interferometric techniques. However, for this parameter choice the signal is entirely dominated by the AQP sector. Indeed, in the limit $G_m\gg G_a$, one finds
\begin{equation}
\Delta_\gamma-\frac{G_m^2\Delta_m+G_a^2\Delta_a}{G_m^2+G_a^2}
\simeq
(\Delta_\gamma-\Delta_m)-\frac{G_a^2}{G_m^2}(\Delta_m-\Delta_a),
\end{equation}
from which the axion-induced correction to the Berry phase follows as
\begin{equation}
\Delta\gamma_a
\simeq
\pi\frac{1}{2G_m}\frac{G_a^2}{G_m^2}(\Delta_a-\Delta_m)
\simeq 2.06\times10^{-34}\pi.
\end{equation}
Thus, for the above benchmark point, the axion contribution is utterly negligible, and the observable geometric phase is effectively an AQP-driven effect.

The benchmark choice considered above lies in case (v), corresponding to the near-resonance regime. This regime is well motivated for power-readout measurements, since resonant enhancement can amplify the observable signal, as emphasized in \cite{SchutteEngel:2021}. For the Berry phase, however, the AQP- and axion-induced contributions add linearly, and for the parameter range of interest the axion contribution is subleading compared with that of the AQP. As a result, any geometric phase observed at the output is expected to be dominated by the AQP contribution. Although one may formally isolate the axion sector by taking the limit $f_\Theta\to \inf$, the corresponding phase generated by an $\mathrm{meV}$-scale axion with $g_{a\gamma\gamma}\sim 10^{-12}~\mathrm{GeV^{-1}}$ remains negligibly small. Hence, to study a potentially observable axion-induced geometric phase, it is more appropriate to eliminate the AQP slab, thereby recovering the setup discussed in Section.~\ref{sec3}.

Besides the branch with $k\simeq k_a$, there also exists a second branch with $k\simeq k_\Theta$ [cf. Eq.~\eqref{soln}]. For this choice, only $\Delta_\gamma$ changes, yielding $|\Delta_\gamma|\simeq 1.69\times 10^{-8}$, which remains much smaller than $|\Delta_m|$ and $|\Delta_a|$. The corresponding hierarchy is therefore $\Delta_a\simeq \Delta_m\gg \Delta_\gamma$. In this regime, the Berry phase may again be analyzed in the bright-dark basis of the $(m,a)$ sector. The axion and AQP sectors are nearly degenerate, while the photon is far detuned from the bright state. The full Berry-phase expression retains the same functional form as in Eq.~\eqref{aqpmain}, but now applies in the detuned regime $\Delta_b\gg \Delta_\gamma$. For $G_m\gg G_a$, the leading axion-induced correction is then given by
\begin{equation}
\Delta\gamma_a\simeq \pm 2\pi(\Delta_\gamma-\Delta_b)\frac{G_a^2}{\left[(\Delta_\gamma-\Delta_b)^2+4G_m^2\right]^{3/2}}.
\end{equation}
This choice of hierarchy results $\Delta\gamma\equiv 0.54\pi$ and $\Delta\gamma_a\simeq 1.75\times 10^{-34}\pi$.

The detection of the axion-induced Berry phase in the presence of AQPs relies on measuring a small phase shift in a coherent THz interferometric signal. The dominant limitation of such measurements arises from noise sources that affect the phase stability rather than the signal power.

In realistic THz spectroscopy setups, technical noise sources dominate over the shot noise. These include phase noise of the THz source, path-length fluctuations due to mechanical vibrations, thermal drifts affecting the refractive index of the medium, and detector readout noise.

The signals for both axion-photon two level and axion-AQP-photon three level systems can be enhanced by extending the effective photon propagation length using multiple reflections within an optical cavity. Since the axion field is insensitive to the mirrors, the geometric phase accumulated by the photons adds coherently over successive reflections. Consequently, after $N$ reflections, the total Berry phase is amplified by a factor of $N$ corresponding to an effective path length $L=Nl$, where $l$ denotes the separation between the mirrors \cite{Raffelt:1987im}.

The phase sensitivity of the interferometric measurement can be further improved by employing squeezed states of light, which reduce the quantum phase uncertainty below the shot-noise limit \cite{Caves:1981hw,LIGOScientific:2011imx,Lang:2013qjs,Eberle:2010zz}. However, such enhancement is only effective when the experiment operates in the quantum noise-limited regime, and may become subdominant in the presence of thermal, technical, or other classical noise sources.

In addition, the sensitivity can be improved by increasing the total integration time, which enables statistical averaging over repeated measurements and reduces the phase uncertainty of the readout \cite{Tam:2011kw}. The experiment needs to be performed at temperature much lower than the frequency to avoid the thermal noise effects \cite{Deppner:2021fks,Christodoulou:2023hyt}.

In the presence of an AQP medium, additional noise can arise from material absorption, dispersion, and finite lifetime of the AQPs modes, which may reduce coherence and introduce amplitude-phase mixing. These effects can be mitigated by operating in a transparency window and by matching the material properties in both interferometer arms.

To enhance the signal-to-noise ratio, it is advantageous to introduce a controlled time dependence in the Berry phase, for example by modulating the external magnetic field or its orientation. This produces a time-dependent phase shift $\gamma(t)$, which imprints a characteristic frequency component in the detected signal. The signal can then be extracted using frequency domain techniques such as FFT or lock-in detection \cite{SYZhang:1991,PMID:18699495}, allowing efficient suppression of broadband noise.

In THz spectroscopy, the detector (e.g., bolometric \cite{2019:NatAs,Lee:2020fen}, heterodyne \cite{Clerk:2008tlb}, or superconducting \cite{Peacock:1996} readout) measures the output intensity of the interferometer, which encodes the phase information through interference. The relevant figure of merit is therefore the phase-equivalent noise rather than the noise-equivalent power. By using matched arms and differential measurements, common-mode noise can be largely canceled, enabling isolation of the axion-induced geometric phase.

\section{Conclusions and outlook}\label{sec5}

In this work, we have studied whether axion-photon interactions can be probed through quantum phases rather than through conventional power- or conversion-based observables. We analyzed two complementary settings: an axion-induced AB phase in a superconducting circuit, relevant when the axion forms a coherent DM background, and an axion-induced Berry phase in interferometric photon propagation through an adiabatically varying magnetic field. We also extended the Berry-phase analysis to a three-level photon-AQP-axion system in a topological medium.

The rf-SQUID proposal provides the strongest phenomenological result of this paper. In the presence of axion DM, the axion-photon interaction induces an effective current and therefore a time dependent magnetic flux through the superconducting loop, which shifts the gauge invariant phase across the JJ and produces a measurable voltage signal. For benchmark parameters, we find a sensitivity around $g_{a\gamma\gamma}^{\mathrm{min}}\sim 7.8\times 10^{-14}~\mathrm{GeV}^{-1}$ at $m_a\sim 10^{-10}~\mathrm{eV}$, with projected reach that can improve on existing limits in that region by roughly one to two orders of magnitude. This identifies AB-phase measurements in superconducting circuits as the most immediately competitive outcome of our analysis and as a promising probe of ultralight axion DM.

The Berry-phase analysis leads to a more qualified conclusion. For the two-level axion-photon system, a MZI with an adiabatically rotating transverse magnetic field defines a clean and conceptually novel geometric phase observable that does not require the axion to constitute DM. However, for conservative tabletop benchmarks we obtain $g_{a\gamma\gamma}^{\mathrm{min}}\sim 6.8\times 10^{-4}~\mathrm{GeV}^{-1}$ at $m_a\sim 2~\mathrm{meV}$, which remains far weaker than current bounds. The main value of this setup is therefore not immediate exclusion power, but rather the identification of a new phase-based observable whose reach could improve if future interferometers achieve substantially better phase stability or exploit quantum-enhanced metrology. In this sense, the Berry-phase proposal should be viewed as a proof-of-principle framework and a longer-term experimental direction.

For completeness, we also evaluated the corresponding Berry phase in the photon-graviton system. The effect is Planck suppressed and, in realistic laboratory conditions, additionally fails the adiabaticity requirement, confirming that the resulting Berry phase is negligible compared with the axion scenarios considered here.

In the three-level photon-AQP-axion system, we find that a Berry phase of order $0.15\pi$ can arise in the THz regime for representative topological material benchmarks. However, the observable phase is dominated by the AQP sector, while the genuine axion-induced correction is negligibly small for the parameter choices considered. This three-level analysis should therefore not be interpreted as a competitive axion-search channel. Its importance lies instead in showing that the same geometric phase formalism remains consistent in a richer coupled system and in identifying a potentially measurable standard physics signal in topological media that could serve as an experimental validation of the framework.

Overall, the main message of this paper is twofold. First, quantum phase observables provide a complementary way to search for axion-photon interactions, with the clearest near-term opportunity coming from the AB-phase measurement in superconducting circuits. Second, geometric phase interferometry remains theoretically well motivated, but its phenomenological impact will depend on future improvements in phase sensitivity, coherence control, and noise suppression. Progress in cryogenic operation, modulation and lock-in strategies, multi-pass geometries, and squeezed or entangled photonic states could substantially sharpen this program. Quantum interferometry therefore offers a useful new tool for axion searches, even though only part of that concept is already competitive with existing constraints.
 
\section*{Acknowledgments}
The authors would like to thank Francesca Chadha-Day for useful discussions. This article is based on the work from COST Actions COSMIC WISPers CA21106 and BridgeQG CA23130, supported by COST (European Cooperation in Science and Technology). T.K.P is supported by the STFC under Grant No. ST/X003167/1.

\section*{Data Availability Statement}
This article has no associated data.

\bibliographystyle{utphys}
\bibliography{reference}
\end{document}